\documentclass[review,authoryear]{elsarticle}

\usepackage{bm,amssymb,amsmath,mathrsfs}
\usepackage{amsthm}
\usepackage{lineno}
\usepackage[usenames]{color}

\usepackage{dcolumn}

\theoremstyle{plain}

\newtheorem*{theorem*}{Theorem}

\renewcommand\thefigure{\arabic{figure}}

\begin{document}

\journal{(internal report CC25-2)}

\begin{frontmatter}

\title{New technique for parameter estimation and improved fits to experimental data for a set of compound Poisson distributions}

\author[cc]{S.~R.~Mane}
\ead{srmane001@gmail.com}
\address[cc]{Convergent Computing Inc., P.~O.~Box 561, Shoreham, NY 11786, USA}

\begin{abstract}
Compound Poisson distributions have been employed by many authors to fit experimental data,
typically via the method of moments or maximum likelihood estimation.
We propose a new technique and apply it to several sets of published data.
It yields better fits than those obtained by the original authors for a set of widely employed compound Poisson distributions (in some cases, significantly better).
The technique employs the power spectrum (the absolute square of the characteristic function).
The new idea is suggested as a useful addition to the tools for parameter estimation of compound Poisson distributions.
\end{abstract}

\begin{keyword}
  Compound Poisson distribution
  \sep parameter estimation
  \sep characteristic function
  \sep power spectrum

\MSC[2020]{
62F30
\sep 60-08
\sep 05-08  
}

\end{keyword}
\end{frontmatter}

\newpage
\section{\label{sec:intro}Introduction}
The monograph by \cite{JohnsonKempKotz} has a comprehensive exposition on compound Poisson distributions.
In most cases, parameter estimation to fit experimental data is accomplished via
standard techniques such as the method of moments or maximum likelihood estimation.
The present paper proposes a new technique for parameter estimation for a widely employed set of compound Poisson distributions.
Fits to several datasets of published experimental data are displayed below;
the new technique yields better fits (in some cases, significantly better) than those obtained the original authors.
It is hoped that the new technique will be a helpful addition to the set of tools to analyze data using the compound Poisson distributions listed below.

\emph{To clarify the point of view of this paper:}
For most of the authors we cite below, their primary interest is a particular set of experimental data,
and statistical distributions are employed as a tool to fit that data.
Our focus is the reverse: we seek to attain a better procedure for parameter estimation (for a set of compound Poisson distributions),
independent of a specific experimental dataset.

For a Poisson process (say `$A$') with rate parameter $\lambda$, the probability generating function (pgf) is $G_A(x) = e^{\lambda(x-1)}$.
Next, consider a `generalizer' distribution $B$, with pgf $G_B(x)$.
The pgf of the resulting compound Poisson distribution is $G(x) = G_A(G_B(x)) = e^{\lambda(G_B(x)-1)}$.
The support of the distribution is $n=0,1,\dots$.
The pgf is expressed as a Maclaurin series in powers of $x$ as follows $G(x) = \sum_{n=0}^\infty P_nx^n$.
Here $P_n$ denotes the probability mass function (pmf).
We shall find it helpful to work with the scaled pmf $h_n=P_n/P_0$. By definition, $h_0=1$.

The monograph by \cite{JohnsonKempKotz} employs the nomenclature by \cite{Gurland1957}, to denote the compounded distribution by $A \vee B$.
The distributions we treat below are
Neyman Type A (Poisson $\vee$ Poisson),
Poisson-Binomial (Poisson $\vee$ binomial),
Poisson-Pascal (Poisson $\vee$ negative binomial) and the
geometric Poisson distribution (Poisson $\vee$ geometric).
See \cite{JohnsonKempKotz}, Secs.~9.6, 9.5, 9.8 and 9.7, respectively.
Their precise definitions will be listed in Sec.~\ref{sec:defs} below.
We shall note some common features of the distributions, which seem to have gone unnoticed in the literature.
Essentially, the histograms of the probability generating functions display a periodic structure,
in a sense to be made precise in Sec.~\ref{sec:modal}.
Motivated by this observation, we suggest that Fourier analysis, i.e.~the use of the characteristic function, might yield a useful tool for parameter estimation.
Recall that the characteristic function is the Fourier transform of the probability generating function.
Specifically, we show how the use of the power spectrum (absolute square of the characteristic function)
furnishes a useful technique for parameter estimation of the above three compound Poisson distributions. 
The basic formalism is developed in Sec.~\ref{sec:Fourierpmf}.
We apply the technique in Sec.~\ref{sec:paramest}, to fit several published datasets,
viz.~\cite{Beall},
\cite{BeallRescia1953} and
\cite{McGBB}.
See also \cite{Gurland1958}, who fitted some of the data of \cite{BeallRescia1953}.
We show how fitting the data using the power spectrum yields better fits than those obtained by the original authors (in some cases, significantly better).
We also fit the data by \cite{FrackerBrischle1944}, using the geometric Poisson distribution.
As a footnote, we also remark on what appear to be computational errors (in some cases) by the original authors,
and show that improved fits are obtained when those errors are corrected.
Sec.~\ref{sec:conc} concludes.

\section{\label{sec:defs}Basic definitions}
\subsection{\label{sec:gen}General remarks}
The material in this section can be found in the monograph by \cite{JohnsonKempKotz}.
However, a submitted paper is required to be self-contained, hence it is necessary to present/explain relevant definitions and parameters below.
All the proofs/derivations of the results in this section are given by \cite{JohnsonKempKotz}.
Explicit expressions for the mean and variance of individual distributions are presented below.
We shall discuss the sample mean and variance when fitting data in Sec.~\ref{sec:paramest}.

\subsection{\label{sec:neyman}Neyman Type A distribution}
The Neyman Type A distribution is obtained by compounding a Poisson distribution with rate $\lambda$ with another Poisson distribution with rate $\phi$.
The mean and variance are (\cite{JohnsonKempKotz}, eq.~(9.122))
\begin{equation}
\label{eq:neyman_mean_var}
\mu = \lambda \phi \,,\qquad
\sigma^2 = \lambda \phi(1+\phi) \,.
\end{equation}

\subsection{\label{sec:binomial}Poisson-binomial distribution}
The Poisson-binomial distribution is obtained by compounding a Poisson distribution with rate $\lambda$ with a binomial distribution with parameters $(k,p)$.
Here $\lambda>0$, $k\in\mathbb{Z}_+$ and $p\in(0,1)$ (and $q=1-p$).
The mean and variance are (\cite{JohnsonKempKotz}, eq.~(9.102))
\begin{equation}
\label{eq:binom_mean_var}
\mu = \lambda kp \,,\qquad
\sigma^2 = \lambda kp(kp +q) \,.
\end{equation}

\subsection{\label{sec:pascal}Poisson-Pascal distribution}
The Poisson-Pascal distribution is obtained by compounding a Poisson distribution with rate $\Lambda$ with a negative binomial distribution with parameters $(k,P)$.
We follow \cite{JohnsonKempKotz} and denote the distribution's parameters by upper case.
Here $\Lambda>0$, $k\in\mathbb{Z}_+$ and $P$ and $Q$ real-valued variables, where $P>0$ and $Q=1+P$.
Note that $P$ and $Q$ are not probabilities.
The mean and variance are (\cite{JohnsonKempKotz}, eq.~(9.159))
\begin{equation}  
\label{eq:pascal_mean_var}
\mu = \Lambda kP \,,\qquad
\sigma^2 = \Lambda kP(kP +Q) \,.
\end{equation}
Observe the formal similarity to eq.~\eqref{eq:binom_mean_var} for the Poisson-binomial distribution.

\subsection{\label{sec:geometric}Geometric Poisson distribution}
The distribution's parameters are a rate $\lambda>0$ and probability $p\in(0,1)$ (and $q=1-p$).
The mean and variance are (\cite{JohnsonKempKotz}, eq.~(9.145))
\begin{equation}
\label{eq:geom_mean_var}
\mu = \frac{\lambda}{q} \,,\qquad
\sigma^2 = \frac{\lambda(1+p)}{q^2} \,.
\end{equation}
The Poisson-Pascal distribution with $k=1$ is equivalent to the geometric Poisson distribution (see \cite{JohnsonKempKotz}).
The relation between the parameters of the two distributions is
$(\lambda,p,q)_{\rm geometric} \leftrightarrow (\Lambda P/Q, P/Q, 1/Q)_{\rm Pascal}$.

\section{\label{sec:modal}Modal structure of selected distributions}
To analyze the shape (histogram) of the pmf, it is better to treat the scaled pmf $h_n=P_n/P_0$, because $h_0=1$ in all cases.
All the graphs in this section treat the scaled pmf $h_n$.
The term `pmf' should be interpreted as `scaled pmf' below.

We begin with Neyman Type A.
The distribution is characterized by two real-valued parameters $\lambda$ and $\phi$, both positive.
We set $\lambda=5$ and $\phi=25$.
The scaled pmf $h_n$ is displayed as the dashed curve in Fig.~\ref{fig:neyman_binom_pascal_pmf_lam5}.
\begin{enumerate}
\item
The dashed curve exhibits multiple local maxima, whose locations are approximately equally spaced at integer multiples of $\phi$, including the origin $n\!\!=\!\!0$.
\item
Anscombe noted this fact in already 1950.
Quoting from \cite{Anscombe} ``In general, the Neyman Type A and Thomas distributions can have any number of modes 
from one upwards, and if there are several modes they will occur at values of $r$ approximately equal to multiples of $m_2$.''
{\em Anscombe defines `mode' to mean `local maximum' and Anscombe's $(r,m_2)$ are our $(n,\phi)$.}
\item
What is also of interest is that the scaled pmfs of the Poisson-binomial and Poisson-Pascal distributions exhibit the same modal structure.
See the other curves in Fig.~\ref{fig:neyman_binom_pascal_pmf_lam5}, viz.~Poisson-binomial (dots) and Poisson-Pascal (dotdash).
For Poisson-binomial, the parameter values are $(\lambda,k,p)=(5,50,0.5)$, i.e.~$kp=25$
and for Poisson-Pascal, they are $(\Lambda,k,P)=(5,50,0.5)$, i.e.~$kP=25$.
\item
For all three distributions, the local maxima occur at values of $n$ which are approximately integer multiples of $\mu/\lambda$.
Observe that $\mu/\lambda=\phi$ for Neyman Type A,
$\mu/\lambda=kp$ for Poisson-binomial and
$\mu/\Lambda=kP$ for Poisson-Pascal, respectively.
\item
For the record, the mean is $\mu=125$ for all three distributions in Fig.~\ref{fig:neyman_binom_pascal_pmf_lam5}.
In all cases, the local maximxa occur at approximately integer multiples of $\mu/\lambda=125/5=25$.
\end{enumerate}
As a sanity check that Fig.~\ref{fig:neyman_binom_pascal_pmf_lam5} is not a special case,
we plot the scaled pmfs of the above distributions for different parameter values in Fig.~\ref{fig:neyman_binom_pascal_pmf_lam10}.
The graphs are for
Neyman Type A (top panel, $(\lambda,\phi)=(10,100)$),
Poisson-binomial (middle panel, $(\lambda,k,p)=(10,1000,0.1)$) and
Poisson-Pascal (bottom panel, $(\Lambda,k,P)=(10,200,0.5)$).
Here $\mu=1000$ and $\lambda=10$ for all three distributions
and the local maximxa occur at approximately integer multiples of $\mu/\lambda=100$.

In summary, it is known that the local maxima of the Neyman Type A distribution are approximately uniformly spaced,
but it does not seem to be recognized that the same holds true for (some) other compound Poisson distributions.
Recall the statements in the introduction about a `generalizer' distribution $B$.
The histogram of the pmf of the above distributions is basically a smooth `background curve' which is `modulated'
with peaks spaced approximately uniformly at integer multiples of $\mathbb{E}[B] = \mu/\lambda$.

\section{\label{sec:Fourierpmf}Power spectrum}
\subsection{Definitions}
The fact that the pmfs of several compound Poisson distributions display periodicities suggests that Fourier analysis of the pmf is helpful.
This is indeed the case, as will be explained below.
The characteristic function $\varphi(t)$ is related to the pgf $G(x)$ via $\varphi(t) = G(e^{it}) = \sum_{n=0}^\infty P_n e^{int}$, i.e.~a Fourier series in $t$.
The power spectrum is its absolute square 
$|\varphi(t)|^2 = \bigl(\sum_{n=0}^\infty P_n \cos(nt)\bigr)^2 +\bigl(\sum_{n=0}^\infty P_n \sin(nt)\bigr)^2$.
For computational work, we evaluate the partial sums for $n=0,\dots,N_{\rm DFT}-1$,
where $N_{\rm DFT}$ is a large integer such that the value of $P_n$ is negligible for $n\ge N_{\rm DFT}$.
(We set $N_{\rm DFT}=2^{10}=1024$ in the studies below.)
Also set $t=2\pi\nu$ where $\nu= j/N_{\rm DFT}$ and $j=0,\dots,N_{\rm DFT}-1$ (hence $\nu\in[0,1)$), and define the partial sums 
\begin{equation}
a(\nu) = \sum_{n=0}^{N_{\rm DFT}-1} P_n \cos(2\pi\nu n) \,, \qquad
b(\nu) = \sum_{n=0}^{N_{\rm DFT}-1} P_n \sin(2\pi\nu n) \,.
\end{equation}
This is a Discrete Fourier Transform (DFT).
There are efficient FFT (Fast Fourier Transform) algorithms to compute the Discrete Fourier Transform.
We refer the reader to the literature on numerical analysis.
Observe that $\lim_{N_{\rm DFT}\to\infty}a(0)=1$, also $b(0)=0$.
For the power spectrum, we compute the partial sum $\Psi(\nu) = (a(\nu))^2 +(b(\nu))^2/a(0)^2$.
Technically, this is a scaled power spectrum, because we normalize $\Psi(0)=1$,
but we shall refer to $\Psi(\nu)$ as the power spectrum below.
We shall see below that the power spectrum is a useful tool for parameter estimation of compound Poisson distributions, to fit experimental data.

\subsection{Aliasing}
We saw in Sec.~\ref{sec:modal} that the histogram of the pmf is `modulated' with peaks spaced approximately uniformly at integer intervals of $\mathbb{E}[B]$,
where $B$ is the `generalizer' distribution.
In Fourier analysis, the graph of the power spectrum of the pmf exhibits a peak at $\nu_{\rm peak}\simeq1/\mathbb{E}[B]$.
The location of that peak provides an estimate of the value of $\mathbb{E}[B]$.
We shall see examples below.

Actually, the above analysis is not completely correct.
Since $0 \le \nu < 1$, the above analysis would imply $\mathbb{E}[B] = 1/\nu_{\rm peak} \ge 1$,
but of course a distribution can have $\mathbb{E}[B] < 1$.
Observe that $\Psi(\nu+m) = \Psi(\nu)$ for $m\in\mathbb{Z}$.
This is because compound Poisson distributions are integer valued distributions, with support $n=0,1,\dots$.
The correct statement is $\mathbb{E}[B] = 1/(\nu_{\rm peak}+m)$, where $m=0,1,\dots$.  
We shall see this below, when fitting experimental data.
Remember that the pmf is only defined at integer values $n=0,1,\dots$.
We cannot sample the pmf in fractional steps $\Delta n = 10^{-3}$, for example.
Our data resolution is constrained to unit steps $\Delta n=1$, which results in aliasing.

\subsection{Example}
Fig.~\ref{fig:neyman_binom_pascal_ps} displays
plots of the power spectra of the Neyman Type A, Poisson-binomial and Poisson-Pascal distributions.
The parameter values are
(i) Neyman Type A $(\lambda,\phi)=(0.4,10)$,
(ii) Poisson-binomial $(\lambda,k,p)=(0.3,20,0.5)$, and
(iii) Poisson-Pascal $(\Lambda,k,P)=(0.5,100,0.1)$.
Hence $\mathbb{E}[B]=\mu/\lambda=10$ for all three distributions.
Observe in Fig.~\ref{fig:neyman_binom_pascal_ps} there is a peak in all three power spectra at $\nu\simeq0.1$.
The location of the peak furnishes an estimate of the value of $\mathbb{E}[B]$ via $\mathbb{E}[B]\simeq1/\nu_{\rm peak}$.
It is also true that there is a mirror peak at $\nu\simeq0.9$, hence the choice of peak is not unique.
It is a theorem in Fourier analysis that the graph of the power spectrum of a real-valued function is symmetric around the midpoint $\nu=\frac12$,
i.e.~$\Psi(\nu) = \Psi(1-\nu)$.
(For the record, $a(\nu) = a(1-\nu)$ and $b(\nu) = -b(1-\nu)$ for a real-valued function.)
Observe also in Fig.~\ref{fig:neyman_binom_pascal_ps} that the power spectrum exhibits a global maximum at $\nu=0$ (aliased with $\nu=1$).
This will cause some complications when fitting experimental data, as we shall see below.

\section{\label{sec:paramest}Parameter estimation \&\ fits to experimental data}
\subsection{Definitions \&\ notation}
\cite{JohnsonKempKotz} have extensive information on parameter estimation for all the distributions listed above (and many others).
I analyzed the experimental data by several authors, viz.~\cite{Beall}, \cite{BeallRescia1953} and \cite{McGBB},
using the power spectrum $\Psi(\nu)$ defined in Sec.~\ref{sec:Fourierpmf}.
I also fitted the data by \cite{FrackerBrischle1944}, using the geometric Poisson distribution.

There are a variety of notations in the literature for parameter estimation, hence we begin with notation.
The experimental data consists of a list of integers, in bins indexed by $n=0,1,\dots$.
Let $c_n$ and $f_n$ respectively denote the observed and fitted counts in each bin.
The value of $c_n$ is an integer but $f_n$ is a real number and depends on the choice of distribution to fit the data.
The value of $c_n$ can be zero, but $f_n$ is always positive for all the distributions we treat.
We denote the total number of experimental observations by $N_c$ (for the `number of counts'), hence $N_c = \sum_{n=0}^\infty c_n$.
By `experimental observations' we mean the set of numbers $x_i$, $i=1,\dots,N_c$, which constitute the raw data.
For example, if $(x_1,\dots,x_{10})=(0,0,0,0,0,2,2,2,3,3)$, then $N_c=10$ and the data lie in four bins and the bin counts are $(c_0,c_1,c_2,c_3)=(5,0,3,2)$.
The sample mean is denoted by $\bar{x}$ and is obtained by summing over the observations as follows.
\begin{equation}
\label{eq:xbar_def}
\bar{x} = \frac{1}{N_c} \sum_{i=1}^{N_c} x_i \,.
\end{equation}
The sample variance is denoted by $s^2$.
\cite{Beall}, \cite{BeallRescia1953} and \cite{FrackerBrischle1944}
computed the sample variance using a denominator of $N_c-1$:
\begin{equation}
\label{eq:sample_var_N1}
s^2_{N_c-1} = \frac{1}{N_c-1}\sum_{i=1}^{N_c} (x_i-\bar{x})^2 \,.
\end{equation}
\cite{McGBB} employed a denominator of $N_c$:
\begin{equation}
\label{eq:sample_var_N}
s^2_{N_c} = \frac{1}{N_c}\sum_{i=1}^{N_c} (x_i-\bar{x})^2 \,.
\end{equation}
The reader should bear this fact in mind.
\begin{enumerate}
\item
We employ a circumflex to denote the estimators computed from the data sample.
\item
For the Neyman Type A distribution, we denote the estimators for $(\lambda,\phi)$ by $(\hat{\lambda},\hat{\phi})$.
For the method of moments, see \cite{JohnsonKempKotz}, eq.~(9.128).
The expressions below are easily derived from eq.~\eqref{eq:neyman_mean_var}.
\begin{equation}
\label{eq:JKK_neyman_est_mm}
\hat{\phi} = \frac{s^2-\bar{x}}{\bar{x}} \,,\qquad
\hat{\lambda} = \frac{\bar{x}}{\hat{\phi}} \,.
\end{equation}
\cite{Beall}, eqs.~(10) and (11),
employed the above formulas, with the sample variance $s^2$ as in eq.~\eqref{eq:sample_var_N1}.
\item
  We shall actually employ the Poisson-Pascal distribution only for the case $k=1$,
  because (some of) the datasets below were fitted using the geometric Poisson distribution
  and we wish to make contact with the fits by those authors.
  For the geometric Poisson distribution,
we denote the estimators for $(\lambda,p)$ by $(\hat{\lambda},\hat{p})$ and also $\hat{q}=1-\hat{p}$.
For the method of moments, see \cite{JohnsonKempKotz}, eq.~(9.149).
The expressions below are easily derived from eq.~\eqref{eq:geom_mean_var}.
\begin{equation}
\label{eq:JKK_geom_est_mm}
\hat{\lambda} = \frac{2\bar{x}^2}{s^2+\bar{x}} \,, \qquad
\hat{p} = \frac{s^2-\bar{x}}{s^2+\bar{x}} \,.
\end{equation}
Alternatively, since $P_0=e^{-\lambda}$ and $h_1=P_1/P_0 = \lambda q$, one can obtain the estimators via
(see \cite{JohnsonKempKotz}, eq.~(9.151))
\begin{equation}
\label{eq:geom_est_P0_h1}
\hat{\lambda} = -\ln(c_0/N_c) \,,\qquad
\hat{q} = \frac{c_1/c_0}{\hat{\lambda}} \,.
\end{equation}
These expressions depend only on the first two counts $c_0$ and $c_1$ (and the total count $N_c$).
We shall eqs.~\eqref{eq:JKK_geom_est_mm} and \eqref{eq:geom_est_P0_h1} when fitting the data by \cite{BeallRescia1953}.
\item
For the Poisson-binomial distribution,
we denote the estimators for $(\lambda,k,p)$ by $(\hat{\lambda},\hat{k},\hat{p})$.
For the method of moments, see \cite{JohnsonKempKotz}, eq.~(9.104).
The expressions below are easily derived from eq.~\eqref{eq:binom_mean_var}.
\begin{equation}
\label{eq:JKK_binom_est_mm}
\hat{p} = \frac{s^2 - \bar{x}}{(k-1)\bar{x}} \,,\qquad
\hat{\lambda} = \frac{\bar{x}}{k\hat{p}} \,.
\end{equation}
\cite{McGBB}, eq.~(8) employed the above formulas, with the sample variance $s^2$ as in eq.~\eqref{eq:sample_var_N}.
\end{enumerate}

\subsection{Goodness of fit}
Recall $c_n$ and $f_n$ respectively denote the observed and fitted counts in each bin $n=0,1,\dots$.
The traditional measure for a goodness of fit is $\chi^2$, defined as follows,
for an integer-valued distribution with support $(0,1,\dots)$,
\begin{equation}
\label{eq:chisqdef}
\chi^2 = \sum_{n=0}^\infty \frac{(c_n-f_n)^2}{f_n} \,.
\end{equation}
However, it is known that $\chi^2$ works well only if the expected count in each bin (the value of $f_n$) is large.
This is not the case for the data in \cite{Beall},
where the printed value of $f_n$ is zero (i.e.~zero to two decimal places, as printed in \cite{Beall})
but the count $c_n$ is one or more.
This contributes a spuriously large value for $(c_n-f_n)^2/f_n$ in eq.~\eqref{eq:chisqdef}.
There are several such examples in Beall's data, e.g.~Table IV in \cite{Beall}.
The conventional solution is to make the bins wider, i.e.~to aggregate the counts in consecutive bins into one wider bin, with a higher count total.
Such a policy works when the author(s) analyze one or a few datasets, which is the case with most papers in the field.
However, it is tedious for a general formalism, to analyze diverse datasets published in different papers.
There is also no uniform policy how to aggregate the bins; it is a personal judgement by each author(s).
We employ instead the following measure for the goodness of fit,
effectively a reduced chi-square statistic per degree of freedom.
Recall $N_c$ is the total count and $s^2$ is the sample variance.
\begin{equation}
\label{eq:Deltadef}
\Delta = \sum_{n=0}^\infty \frac{(c_n-f_n)^2}{N_cs^2} \,.
\end{equation}
Employing $\Delta$ has the merit that aggregating the bins is unnecessary.
One can write an automated computer program to compute $\Delta$, for multiple datasets in papers by different authors.
I append a subscript such as $\Delta_{\rm MM}$ or $\Delta_{\rm PS}$ to denote the value of $\Delta$ computed by using the method of moments or the power spectrum, etc.

\subsection{Beall (1940): general remarks}
There are four tables in \cite{Beall}.
The data in Table III are the same as in Table II (Beall simply presented additional analysis), hence we treat the data in Tables I, II and IV below.
The sample mean and variance were computed using eqs.~\eqref{eq:xbar_def} and \eqref{eq:sample_var_N1}, respectively.
Beall employed the Neyman Type A distribution to fit the data and denoted the estimators for $(\lambda,\phi)$ by $(m_1,m_2)$.
Beall employed the method of moments and calculated the values of $m_1$ and $m_2$
from the sample mean and variance using eq.~\eqref{eq:JKK_neyman_est_mm} (see \cite{Beall}, eqs.~(10) and (11)).
In all cases, the goodness of fits are denoted by $\Delta_{\rm MM}$ using the method of moments and $\Delta_{\rm PS}$ using the power spectrum.

\subsection{Beall Table I}
\cite{Beall} found the data in Table I were well fitted by the Neyman Type A distribution.
There are four columns in Table I.
The total count was $N_c=325$ in every column.
For the data in Column 1,
the sample mean and variance are $\bar{x} = 1.4$ and $s^2 \simeq 2.3272$.
From this, I obtained the estimated parameter values $m_1 \simeq 2.1140$ and $m_2 \simeq 0.6623$, using eq.~\eqref{eq:JKK_neyman_est_mm}.
Beall did not publish values for $m_1$ and $m_2$, but using the numbers above,
my fitted values for the counts in each bin matched Beall's published numbers closely.
{\em The same was true for the data in all the other columns,
  which suggests that my computed values for $m_1$ and $m_2$ match what Beall obtained.}
I have little to add to Beall's analysis.
Overall, the data in each column of Table I were well fitted by the Neyman Type A distribution, using the method of moments.
My results for the estimators $(m_1,m_2)$ and the goodness of fit $\Delta$ are tabulated in Table \ref{tb:Beall_TableI_fit}.

\subsection{Beall Table II}
\subsubsection{General remarks}
There are four columns in Table II.
The total count was $N_c=120$ in every column.
For every column, the count total, values of the sample mean and variance and the estimates for $m_1$ and $m_2$
(using the method of moments) are listed in Table \ref{tb:Beall_Table2}.
(Beall did not publish values for $m_1$ and $m_2$.)
The estimator values $\hat{\lambda}$ and $\hat{\phi}$ using my fits, and the goodness of fit (method of moments and myself), are also tabulated.
Beall reported poor fits for the data in Table II (except for column 2), using the Neyman Type A distribution.
However, I was able to obtain improved fits, which were reasonably good for every column.
To do so, I made extensive use of the power spectrum of the pmf (see Sec.~\ref{sec:Fourierpmf}).
I also encountered issues of aliasing, which will be explained below.

\subsubsection{Column 1}
Beall claimed to obtain a poor fit to the data.
I employed the power spectrum to fit the data as follows.
The power spectrum is plotted in Fig.~\ref{fig:Beall_TableII_Col1_ps} (upper panel),
with a magnified view in the lower panel.
There is a local maximum at $\nu=0.42$, hence I tried $\hat{\phi}=1/0.42\simeq2.38$,
but I obtained a better fit using the mirror peak $\hat{\phi}=1/0.58\simeq1.724$.
The value of $\hat{\lambda}$ was computed using $\hat{\lambda}=\bar{x}/\hat{\phi} = 2.3393$.
The above values and the goodness of fits are tabulated in the first row of Table \ref{tb:Beall_Table2}.
The improvement in the fit, using the power spectrum, is significant.
The fits are displayed in the top panel of Fig.~\ref{fig:Beall_TableII_Col12_fits}.
The points denote the count $c_n$ in each bin and
the dashed and dotdash curves are the fits using the method of moments and power spectrum, respectively.
The data points vaguely exhibit local maxima spaced at $1.724$ (local maxima at $n=0,2,3,5,7,9$).
One can see visually that the dotdash curve (using the power spectrum) yields a better fit.
Beall's fit grossly overestimates the value at $n=0$ (fit $= 34.4$, data $=19$)
and grossly underestimates the value at $n=1$ (fit $= 6.4$, data $=12$).
The fit using the power spectrum does much better: 
($n=0$, fit $= 17.6$, data $=19$) and
($n=1$, fit $= 12.6$, data $=12$)
and it also does a much better job for the points at $n=2$ and $3$.
For $n\ge4$ the fits by Beall and I match the data about equally well.

\subsubsection{Column 2}
Beall claimed to obtain a good fit to the data.
I employed the power spectrum to fit the data as follows.
The power spectrum is plotted in Fig.~\ref{fig:Beall_TableII_Col2_ps} (upper panel),
with a magnified view in the lower panel.
There is a local maximum at $\nu=0.42$, hence I tried $\hat{\phi}=1/0.32=3.125$,
but I obtained a better fit using the mirror peak $\hat{\phi}=1/0.68\simeq1.4706$.
The value of $\hat{\lambda}$ was computed using
$\hat{\lambda}=\bar{x}/\hat{\phi} = 2.1533$.
The above values and the goodness of fits are tabulated in the second row of Table \ref{tb:Beall_Table2}.
They are approximately equal to those obtained using the method of moments.
The fits are displayed in the bottom panel of Fig.~\ref{fig:Beall_TableII_Col12_fits}.
The points denote the count $c_n$ in each bin and
the dashed and dotdash curves (visually indistinguishable) are the fits using the method of moments and power spectrum, respectively.
Beall obtained a good fit, and I achieved only a small improvement.

\subsubsection{Column 3}
Beall claimed to obtain a poor fit to the data.
I employed the power spectrum to fit the data as follows.
The power spectrum is plotted in the top panel Fig.~\ref{fig:Beall_TableII_Col34_ps}.
I puzzled over the power spectrum until I realized that the (almost) local maximum
(more likely an inflection point) at $\nu=0.25$
is aliased and the correct value to use is $1.25$, hence $\hat{\phi}=1/1.25=0.8$.
The data is displayed in the top panel of Fig.~\ref{fig:Beall_TableII_Col34_fits}.
It does not exhibit local maxima spaced at $1/0.25=4$, but is unimodal,
which suggests that the spacing between `local maxima' is \emph{less than unity}, i.e.~$\hat{\phi}<1$, i.e.~aliasing.
The value of $\hat{\lambda}$ was computed using
$\hat{\lambda}=\bar{x}/\hat{\phi} \simeq 1.8542$.
The above values and the goodness of fits are tabulated in the third row of Table \ref{tb:Beall_Table2}.
The fits are displayed in the top panel of Fig.~\ref{fig:Beall_TableII_Col34_fits}.
There is a moderate but noticeable improvement using the power spectrum.

\subsubsection{Column 4}
Beall claimed to obtain a poor fit to the data.
I employed the power spectrum to fit the data as follows.
The power spectrum is plotted in the bottom panel Fig.~\ref{fig:Beall_TableII_Col34_ps}.
As with the power spectrum for column 3, I puzzled over this.
There is an approximate local maximum at $\nu=0.75$,
until I realized the best choice is the global maximum itself at $\nu=1$,
hence $\hat{\phi}=1/1=1$.
The value of $\hat{\lambda}$ was computed using
$\hat{\lambda}=\bar{x}/\hat{\phi} \simeq 1.5083$.
The above values and the goodness of fits are tabulated in the fourth row of Table \ref{tb:Beall_Table2}.
The fits are displayed in the bottom panel of Fig.~\ref{fig:Beall_TableII_Col34_fits}.
There is a moderate but noticeable improvement using the power spectrum.

\subsubsection{Beall Table II: summary}
\begin{enumerate}
\item
  {\em In fairness, Beall's paper was published in 1940, when there was no access to modern computing machinery, to perform the computations I was able to do.}
\item
  By examining the peak structure of the power spectrum of the pmf,
  I was able to obtain better fits to Beall's data.
  Nevertheless, personal judgement was required to analyze the peaks in the power spectrum, including problems of aliasing.
\item
  Beall reported the fits in columns 1, 3 and 4 to be poor.
  In all three cases, Beall's fits overestimated the value at $n=0$ and
  underestimated the value at $n=1$, but for $n>1$ it seems to me Beall's fits were not too bad.
  I obtained significant improvement for column 1 and moderate improvements for columns 3 and 4.
  In all cases, using the power spectrum fitted the data much better at $n=0$ and $1$,
  but were not too different from Beall for $n>1$.  
\item
  Beall stated that improved (reasonably good) fits were obtained by smoothing the data.
I did not find it necessary to smooth the data.
\end{enumerate}

\subsection{Beall Table IV}
\subsubsection{General remarks}
There are three columns in Table IV.
Unlike the previous tables, the count total $N_c$ in Table IV is different for each column, viz.~$N_c=31$, $67$ and $70$, respectively.
As with the other tables, Beall computed the values of $m_1$ and $m_2$ using the method of moments.
However, a peculiar feature of Table IV is that when I applied the method of moments to fit the data,
my fitted values for the counts in each bin looked nothing like Beall's fitted numbers for columns 1 and 2.
My fits, using the method of moments, are much better than Beall's.
Several of Beall's fitted values in columns 1 and 2 are zero, which is impossible:
it really means the numbers are zero when rounded to two decimal places.
It was only for column 3 that my numbers matched Beall's fitted values.
This leads me to suspect that Beall made a computational error in columns 1 and 2.
For every column, the count total, values of the sample mean and variance and the estimates for $m_1$ and $m_2$
(using the method of moments) are listed in Table \ref{tb:Beall_Table4}.
(Beall did not publish values for $m_1$ and $m_2$.)
The estimator values $\hat{\lambda}$ and $\hat{\phi}$ using my fits, and the goodness of fit (method of moments and myself), are also tabulated.

\subsubsection{Column 1}
My fitted values using the method of moments are very different from Beall's.
I employed the power spectrum to fit the data as follows.
The power spectrum is plotted in the top panel of Fig.~\ref{fig:Beall_TableIV_Col123_ps}.
At first glance, it looks like noise but let us not rush to judgement.
A magnified view of the power spectrum is plotted in Fig.~\ref{fig:Beall_TableIV_Col1_mag_ps}.
Several peaks are visible. They are admittedly not tall, but remember it is their {\em locations} which matter.
We loop through them, starting with the peak at $\nu\simeq0.093$,
and compute the value of $\Delta$ by setting $\phi=1/\nu_{\rm peak}$ for each peak (and $\lambda = \bar{x}/\phi$).
The results are tabulated in Table \ref{tb:Beall_TableIV_Col1_Delta}.
The peak at $\nu=0.215$ yields the smallest value of $\Delta$.
Hence we set $\hat{\phi}=1/0.215\simeq4.6512$ and $\hat{\lambda}=\bar{x}/\hat{\phi}\simeq3.7105$.
The above values and the goodness of fits are tabulated in the first row of Table \ref{tb:Beall_Table4}.
The fits are displayed in the top panel of Fig.~\ref{fig:Beall_TableIV_Col123_fits}.
There is too much scatter in the data to claim there are local maxima spaced at approximately $4.65$.
The curve using the method of moments is my calculation, and does not match Beall.
Observe that the estimated values for $\lambda$ and $\phi$ are noticeably different, using the method of moments and the power spectrum,
but the fitted curves are relatively close to each other and the goodness of the fits is about equal.
There is so much scatter in the data that there is probably a large set of parameter values which all yield fits close to the best.
{\em In other words, the method of moments yields a satisfactory fit to the data.}
I suspect Beall simply made a computational error.

\subsubsection{Column 2}
My fitted values using the method of moments are very different from Beall's.
I employed the power spectrum to fit the data as follows.
The power spectrum is plotted in the middle panel of Fig.~\ref{fig:Beall_TableIV_Col123_ps}.
There is a discernible local maximum at $\nu=0.27$,
yielding the estimated values $\hat{\phi}=1/0.27\simeq3.7037$ and $\hat{\lambda}=\bar{x}/\hat{\phi}\simeq1.1525$.
The above values and the goodness of fits are tabulated in the second row of Table \ref{tb:Beall_Table4}.
The fits are displayed in the middle panel of Fig.~\ref{fig:Beall_TableIV_Col123_fits}.
The data indeed vaguely exhibit local maxima spaced at $3.7$ (local maxima at $n=0,4,8,11,15$).
The curve using the method of moments is my calculation, and does not match Beall.
Fitting the data using the power spectrum yields a significantly better result than using the method of moments.

\subsubsection{Column 3}
In this case, my fitted values using the method of moments are close to Beall's numbers.
I employed the power spectrum to fit the data as follows.
The power spectrum is plotted in the bottom panel of Fig.~\ref{fig:Beall_TableIV_Col123_ps}.
There is a local maximum at $\nu=0.71$, 
yielding the estimated values $\hat{\phi}\simeq1/0.71\simeq1.4085$ and $\hat{\lambda}=\bar{x}/\hat{\phi}\simeq1.5214$.
The above values and the goodness of fits are tabulated in the third row of Table \ref{tb:Beall_Table4}.
The fits are displayed in the bottom panel of Fig.~\ref{fig:Beall_TableIV_Col123_fits}.
In this case, the curve using the method of moments matched Beall's numbers.
Fitting the data using the power spectrum yields a significantly better result than using the method of moments.

\subsubsection{Beall Table IV: summary}
\begin{enumerate}
\item
  For Table IV, the term `method of moments' refers to my fits, not Beall's.
\item
  For column 1, there is so much scatter in the data that one can find many different sets of parameter values which fit the data equally well.
  The method of moments yielded a satisfactory fit.
\item
  For columns 2 and 3, the power spectrum exhibited a discernible peak,
  and the resulting fit was much better than using the method of moments.
\end{enumerate}

\subsection{Beall \&\ Rescia (1953) and Gurland (1958)}
\cite{BeallRescia1953} published several tables of data.
They fitted their data using a distribution indexed by a parameter $n=0,1,\dots$,
where $n=0$ is equivalent to Neyman Type A.
\cite{Gurland1958} showed that the limit $n\to\infty$
is equivalent to the geometric Poisson distribution.
I analyze only the two tables with the largest number of populated bins, viz.~Tables V and VIII.

{\em Fitting of data:}
We treat only the fits using $n=0$ and $n\to\infty$.
The latter case is equivalent to geometric Poisson.
Let us discuss how to fit this.
Observe that the periodicites in the pmf displayed in Sec.~\ref{sec:modal} were for Neyman Type A, Poisson-binomial and Poisson-Pascal.
The geometric Poisson distribution is equivalent to Poisson-Pascal with $k=1$.
Hence we fit the data using Poisson-Pascal with $k=1$. See below.
{\em Notation for fits:} `NTA' = Neyman Type A, `geom' = geometric Poisson, `PP' = Poisson-Pascal, `MM' = method of moments and `PS' = power spectrum.

\subsubsection{Beall \&\ Rescia Table V}
The total count is $N_c=7640$.
The sample mean and variance (using eq.~\eqref{eq:sample_var_N1})
are $\bar{x}\simeq 0.1068$ and $s^2 \simeq 0.2944$.
We begin with the fit using the case $n=0$ (Neyman Type A).
My computed fit using the method of moments (eq.~\eqref{eq:JKK_neyman_est_mm})
approximately matched Beall and Rescia's published numbers.
I employed the power spectrum to fit the data as follows.
The power spectrum is displayed in Fig.~\ref{fig:BR_Table5_PS1000}.
Select the global maximum at $\nu=1$. This yields $\hat{\phi}=1/1=1$
and $\hat{\lambda}=\bar{x}/\hat{\phi}\simeq 0.1068$.
The computed fits are tabulated in the first and second rows of Table.~\ref{tb:BR_Delta}.
Using the power spectrum yields a much better fit than the method of moments.

Next we fit the data using the case $n\to\infty$ (geometric Poisson).
My computed fit using the method of moments (eq.~\eqref{eq:JKK_geom_est_mm})
approximately matched Beall and Rescia's published numbers.
The alternative fit using eq.~\eqref{eq:geom_est_P0_h1} yields a much better fit.
I employed the power spectrum to fit the data as follows.
Recall from above that Poisson-Pascal with $k=1$ was actually employed to fit the data.
See Secs.~\ref{sec:pascal} and \ref{sec:geometric},
also eqs.~\eqref{eq:pascal_mean_var} and \eqref{eq:geom_mean_var}.
We equate $1/(kP)$, i.e.~$1/P$, to the location of a peak $\nu_{\rm peak}$ in the power spectrum.
In this case, the best choice is the local maximum at $\nu=0.5$.
This is aliased and the correct value to employ is $\nu+1=1.5$,
whence the estimator value is $\hat{P} = 1/1.5=\frac23$ and $\hat{Q}=1+\hat{P}=\frac53$.
Next $\hat{\Lambda} = \bar{x}/(k\hat{P}) \simeq 0.1602$.
We compute the fit using Poisson-Pascal with the above inputs.
The computed fits are tabulated in the third, fourth and fifth rows of Table.~\ref{tb:BR_Delta}.
Using the power spectrum yields a slightly better fit than the alternative fit using eq.~\eqref{eq:geom_est_P0_h1}
and both are significantly better than using the method of moments.

The fits (using the power spectrum) are displayed in the top panel of Fig.~\ref{fig:BR_Table58_fits}
(dash = Neyman Type A and dotdash = Poisson-Pascal).
The point at $n=0$ has the value $7178$ and was excluded because it is much taller than the others, which are all less than $300$.
One can see that using Poisson-Pascal yields a better fit.

\subsubsection{Beall \&\ Rescia Table VIII}
The total count is $N_c=772$.
The sample mean and variance (using eq.~\eqref{eq:sample_var_N1})
are $\bar{x}\simeq 2.4016$ and $s^2 \simeq 7.0941$.
We begin with the fit using the case $n=0$ (Neyman Type A).
My computed fit using the method of moments (eq.~\eqref{eq:JKK_neyman_est_mm})
approximately matched Beall and Rescia's published numbers.
I employed the power spectrum to fit the data as follows.
The power spectrum is displayed in Fig.~\ref{fig:BR_Table8_PS1000}.
There is a local maximum at $\nu\simeq 0.69$.
This yields $\hat{\phi}=1/0.69\simeq 1.4493$
and $\hat{\lambda}=\bar{x}/\hat{\phi}\simeq 1.6571$.
The computed fits are tabulated in the first and second rows of Table.~\ref{tb:BR_Delta}.
Using the power spectrum yields a much better fit than the method of moments.

Next we fit the data using the case $n\to\infty$ (geometric Poisson).
\cite{Gurland1958} also fitted Table VIII, using geometric Poisson,
and obtained the same fitted numbers published by Beall and Rescia.
My computed fit using the method of moments (eq.~\eqref{eq:JKK_geom_est_mm})
approximately matched Beall and Rescia's published numbers.
The alternative fit using eq.~\eqref{eq:geom_est_P0_h1} yields a better fit.
I employed the power spectrum to fit the data as follows.
Recall from above that Poisson-Pascal with $k=1$ was employed.
In this case, the best choice is the local maximum at $\nu\simeq 0.31$.
This is aliased and the correct value to employ is $\nu+1=1.31$,
whence the estimator value is $\hat{P} = 1/1.31\simeq 0.7634$ and $\hat{Q}=1+\hat{P}\simeq 1.7634$.
Next $\hat{\Lambda} = \bar{x}/(k\hat{P}) \simeq 3.1460$.
We compute the fit using Poisson-Pascal with the above inputs.
The computed fits are tabulated in the third, fourth and fifth rows of Table.~\ref{tb:BR_Delta}.
Using the power spectrum and the alternative fit using eq.~\eqref{eq:geom_est_P0_h1}
yield approximately equal results and both are slightly better than using the method of moments.

The fits (using the power spectrum) are displayed in the bottom panel of Fig.~\ref{fig:BR_Table58_fits}
(dash = Neyman Type A and dotdash = Poisson-Pascal).
The fit using Neyman Type A exhibits a shoulder at $n=1$ and $2$.
It is trying to exhibit a local maximum at $\phi\simeq1.5$, but this does not fit the data well.
The fit using Poisson-Pascal is better.

\subsubsection{Summary}
For both Tables V and VIII, fitting the data using geometric Poisson was (much?) better than using Neyman Type A.
The use of the power spectrum, for both Neyman Type A and geometric Poisson
(more accurately, Poisson-Pascal with $k=1$) is competitive, or better,
than the fits obtained using traditional techniques.
Admittedly, some judgement is required to select peaks and treat aliasing.
In the case of geometric Poisson, the use of eq.~\eqref{eq:geom_est_P0_h1},
as opposed to the method of moments (eq.~\eqref{eq:JKK_geom_est_mm}), is also a good choice.

\subsection{McGuire, Brindley and Bancroft (1957)}
\subsubsection{General remarks}
\cite{McGBB} published nine tables of data.
They fitted their data using (i) negative binomial, (ii) Neyman Type A, and (ii) Poisson-binomial.
They employed eq.~\eqref{eq:sample_var_N} to compute the sample variance.
We select the two datasets with the largest number of populated bins, viz.~Distributions 3 and 5.
They yield histograms with the most structure.
I employed the method of moments,
using eq.~\eqref{eq:JKK_neyman_est_mm} for Neyman Type A and
eq.~\eqref{eq:JKK_binom_est_mm} for Poisson-binomial.
I briefly treat the negative binomial distribution at the end,
where they stated that it gave the best fit for Distribution 5.
{\em Notation for fits:} `PB' = Poisson-binomial, `NTA' = Neyman Type A, `NB' = negative binomial,
`MM' = method of moments and `PS' = power spectrum.

\subsubsection{Distribution 3}
\cite{McGBB} stated that the total count for Distribution 3 is $N_c=311$, but I summed their counts and obtained $N_c=324$.
Using $N_c=324$, I obtained $\bar{x} \simeq 25.633$ and $s^2 \simeq 82.368$ for the sample mean and variance,
which agrees with their numbers to three decimal places (\cite{McGBB} Table 1).
Using the method of moments, they stated that they obtained the best fit using the Poisson-binomial distribution (they set $k=4$), and Neyman Type A was second best.
I found this to be so (just slightly better).
I employed the power spectrum to fit the data as follows.
The power spectrum (magnified view) is plotted in the top panel of Fig.~\ref{fig:McGBB_dist35_PS1000}.
There is a peak at $\nu\simeq0.46$.
This yields the estimated value $\hat{\phi}=1/0.46\simeq2.1739$ for Neyman Type A
and $(kp)_{\rm est}=1/0.46$ for Poisson binomial.
Setting $k=4$ yields the estimated value $\hat{p}\simeq0.5435$.
In both cases, the estimated value of $\lambda$ is
$\hat{\lambda} = \bar{x}/\hat{\phi} \simeq 11.7910$.
The goodness of fits of all the methods are tabulated in the third column of Table \ref{tb:McGBB_Delta}.
Fitting the data using the power spectrum yielded
a slightly worse fit than the method of moments for Poisson-binomial, but was about equal for Neyman Type A.
A quick check also reveals that $k=4$ is the optimal choice for Poisson-binomial.
Fig.~\ref{fig:McGBB_dist35_hist} (top panel)
displays the data points and the fit using Neyman Type A, computed using the power spectrum.
There is considerable scatter in the data.
This likely has the consequence that multiple distributions may fit the data equally well.

\subsubsection{Distribution 5}
\cite{McGBB} stated that the total count for Distribution 5 is $N_c=324$, which I confirmed to be correct.
I computed the sample mean and variance to be $\bar{x} \simeq 5.231$ and $s^2 \simeq 10.740$.
{\em These values disagree with \cite{McGBB}, Table 1,
  which are $\bar{x} = 5.228$ and $s^2 = 10.621$.
I do not know why.}
Using the method of moments, they stated that
Neyman Type A fitted the data better than Poisson-binomial (for this distribution they set $k=3$).
I found this to be so, but I employed my values for the sample mean and variance.
They also stated that the negative binomial distribution gave the best fit of all.
I treat the negative binomial distribution briefly at the end.
I employed the power spectrum to fit the data as follows.
The power spectrum (magnified view) is plotted in the bottom panel of Fig.~\ref{fig:McGBB_dist35_PS1000}.
There is a peak at $\nu\simeq0.26$ but this is aliased and the correct value is $1.26$.
This yields the estimated value $\hat{\phi}=1/1.26\simeq0.7937$ for Neyman Type A
and $(kp)_{\rm est}=1/1.26$ for Poisson binomial.
Setting $k=3$ yields the estimated value $\hat{p}\simeq0.2646$.
In both cases, the estimated value of $\lambda$ is
$\hat{\lambda} = \bar{x}\times1.26 \simeq 6.5917$.
The goodness of fits of all the methods are tabulated in the fourth column of Table \ref{tb:McGBB_Delta}.
Fitting the data using the power spectrum yielded noticeably better results than the method of moments.
A quick check also reveals that $k=3$ is {\em not} the optimal choice for Poisson-binomial.
Better is $k\to\infty$ (with $kp$ fixed), i.e.~Neyman Type A.
Using $k=1000$ yields a fit indistinguishable from Neyman Type A.
Fig.~\ref{fig:McGBB_dist35_hist} (bottom panel)
displays the data points and the fit using Neyman Type A, computed using the power spectrum.
There is an outlier point at $n=5$,
which almost certainly dominates the contribution to the goodness of fit and
complicates fitting the data by any distribution.

\subsubsection{Negative binomial distribution}
\cite{McGBB} stated that the
negative binomial distribution gave the best fit for Distribution 5.
Curiously, they just said `negative binomial' but did not specify the value of $n$ (or $k$ in my notation).
They cited \cite{BlissFisher1953} for the fitting procedure for the negative binomial distribution.
For the record, I fitted the data in Distribution 5 using the negative binomial distribution (`NB') and the method of moments.
For the negative binomial distribution with parameters $(k,p)$,
with $k\in\mathbb{R}_+$, $p\in(0,1)$ and $q=1-p$, the mean and variance are
\begin{equation}
\label{eq:negbinom_mean_var}
\mu = \frac{kq}{p} \,,\qquad
\sigma^2 = \frac{kq}{p^2} \,.
\end{equation}
The estimators $(\hat{k},\hat{p})$ for $(k,p)$ are given below, where the formula for $\hat{k}$ is from \cite{BlissFisher1953}, eq.~(3).
The expressions below are easily derived from eq.~\eqref{eq:negbinom_mean_var}.
\begin{equation}
\hat{k} = \frac{\bar{x}^2}{s^2-\bar{x}} \,, \qquad
\hat{p} = \frac{\hat{k}}{\hat{k}+\bar{x}} \,.
\end{equation}
Recall $\bar{x}\simeq5.231$ and $s^2\simeq10.740$ are the values of the sample mean and variance.
This yields $\hat{k}\simeq4.97$.
Actually, the value of $k$ does not have to be an integer for the negative binomial distribution (if the support is $n=0,1,\dots$),
but I set $\hat{k}=5$ and computed $\hat{p}=5/(5+\bar{x})\simeq0.4887$.
The resulting goodness of fit was $\Delta_{\rm NB} \simeq 128.5 \times 10^{-3}$, tabulated in the last row of Table \ref{tb:McGBB_Delta}.
Comparing to the other numbers in Table \ref{tb:McGBB_Delta}, this is indeed a better fit than using the method of moments
for Neyman Type A or Poisson-binomial (with $k=3$).
However, it is not as good as the fits obtained using the power spectrum.
Fig.~\ref{fig:McGBB_dist5_hist} displays a plot the fit,
using Neyman Type A (dotdash, same as the bottom panel in Fig.~\ref{fig:McGBB_dist35_hist})
and the negative binomial (dashed curve), with the above parameter values.

I ran a scan for integer values of $k$ and set $p=k/(k+\bar{x})$ and the optimal choice, i.e.~smallest value of $\Delta$, was given by $(k,p)=(7,0.5723)$.
This yielded $\Delta \simeq 110.1 \times 10^{-3}$, which is about equal to the fits using the power spectrum.

\subsubsection{Summary}
This lends credibility to the claim that fitting the data using the power spectrum is competitive with traditional fitting techniques.
It also shows that all three distributions
(Poisson-binomial, Neyman Type A, negative binomial) are capable of yielding approximately equally good fits to the data in Distribution 5.
{\em In other words, there is no clear choice of the best distribution to fit the data.}

For the record, I also fitted the data in Distribution 3 using the negative binomial distribution.
I obtained $\hat{k}\simeq11.58$ and this time I left it as fractional and did not round to an integer.
Then $\hat{p}=\hat{k}/(\hat{k}+\bar{x})\simeq0.3112$ and $\Delta_{\rm NB} \simeq 8.99 \times 10^{-3}$, tabulated in the last row of Table \ref{tb:McGBB_Delta}.
In agreement with \cite{McGBB}, the ranking of the best fits is (Poisson-binomial, Neyman Type A, negative binomial);
see the third column of Table \ref{tb:McGBB_Delta}.
Running a scan of integer values of $k$ and setting $p=k/(k+\bar{x})$ yielded that $k=11$ or $k=12$ are indeed the best choices.

\subsection{Fracker and Brischle (1944)}
\cite{FrackerBrischle1944} fitted multiple sets of data using the Poisson distribution and the Neyman Type A distribution
(which they called the ``contagious distribution'').
Let us denote Neyman Type A by ``NTA'' for brevity.
We fit five of their datasets below (Table II: Kaniksu, Mt.~Spokane, Clearwater and Table III: Beaver Creek and Cow Creek).
They stated that neither Poisson nor NTA fitted their data well.
They obtained the best fit by using a linear combination of the two distributions, viz.~($\frac13\times$Poisson +$\frac23\times$NTA).
(Note: for Cow Creek they used ($0.4\times$Poisson +$0.6\times$NTA).)
We confirmed their fits numerically.
\emph{However, we obtained much better fits using the geometric Poisson distribution.}
Using eq.~\eqref{eq:geom_est_P0_h1} yielded better fits than the method of moments eq.~\eqref{eq:JKK_geom_est_mm}.
Our values for the estimators $(\hat{\lambda},\hat{p})$ for the datasets,
calculated using eq.~\eqref{eq:geom_est_P0_h1},
are tabulated in Table \ref{tb:FB_fits}.
The resulting fits to the data are displayed in Fig.~\ref{fig:FB_hist}.
The power spectrum was not employed in this section.
The main conclusion is that the datasets of \cite{FrackerBrischle1944} are really fitted by the geometric Poisson distribution,
not a linear combination of Poisson and Neyman Type A.
One can observe visually that in all five datasets in Fig.~\ref{fig:FB_hist},
the geometric Poisson distribution fits the data almost perfectly.
There is some experimental scatter in the last dataset (Cow Creek),
but the fitted curve interpolates between the scatter, which is the best that can be expected.

\section{\label{sec:conc}Conclusion}
As stated in the Introduction,
for most of the authors we have cited, their primary interest is a particular set of experimental data,
and statistical distributions are employed as a tool to fit that data.
Our focus is the reverse: we seek to attain a better procedure for parameter estimation (for a set of compound Poisson distributions),
independent of a specific experimental dataset.
Specifically, we suggest a new technique for parameter estimation, employing the power spectrum, for a set of compound Poisson distributions
(Neyman Type A, Poisson-binomial and Poisson-Pascal).
The power spectrum was employed in Sec.~\ref{sec:paramest} to fit experimental datasets published by several authors.
It yielded competitive or better fits than those obtained by the original authors.
We not only computed goodness-of-fit numbers, but plotted several graphs to demonstrate the better quality of our fits to the data.

Having said the above, one must acknowledge that the use of the power spectrum for parameter estimation is still in its infancy.
It was tested mostly with the Neyman Type A distribution, using data from multiple authors,
and a few examples using the Poisson-Pascal and Poisson-binomial distributions.
Although the new technique performed well for the cases studied in this paper, additional testing is required to validate its merits.
It is a matter for future research.


\setcounter{figure}{0}
\renewcommand\thefigure{\arabic{figure}}
\newpage
\begin{figure}[!htb]
\centering
\includegraphics[width=0.75\textwidth]{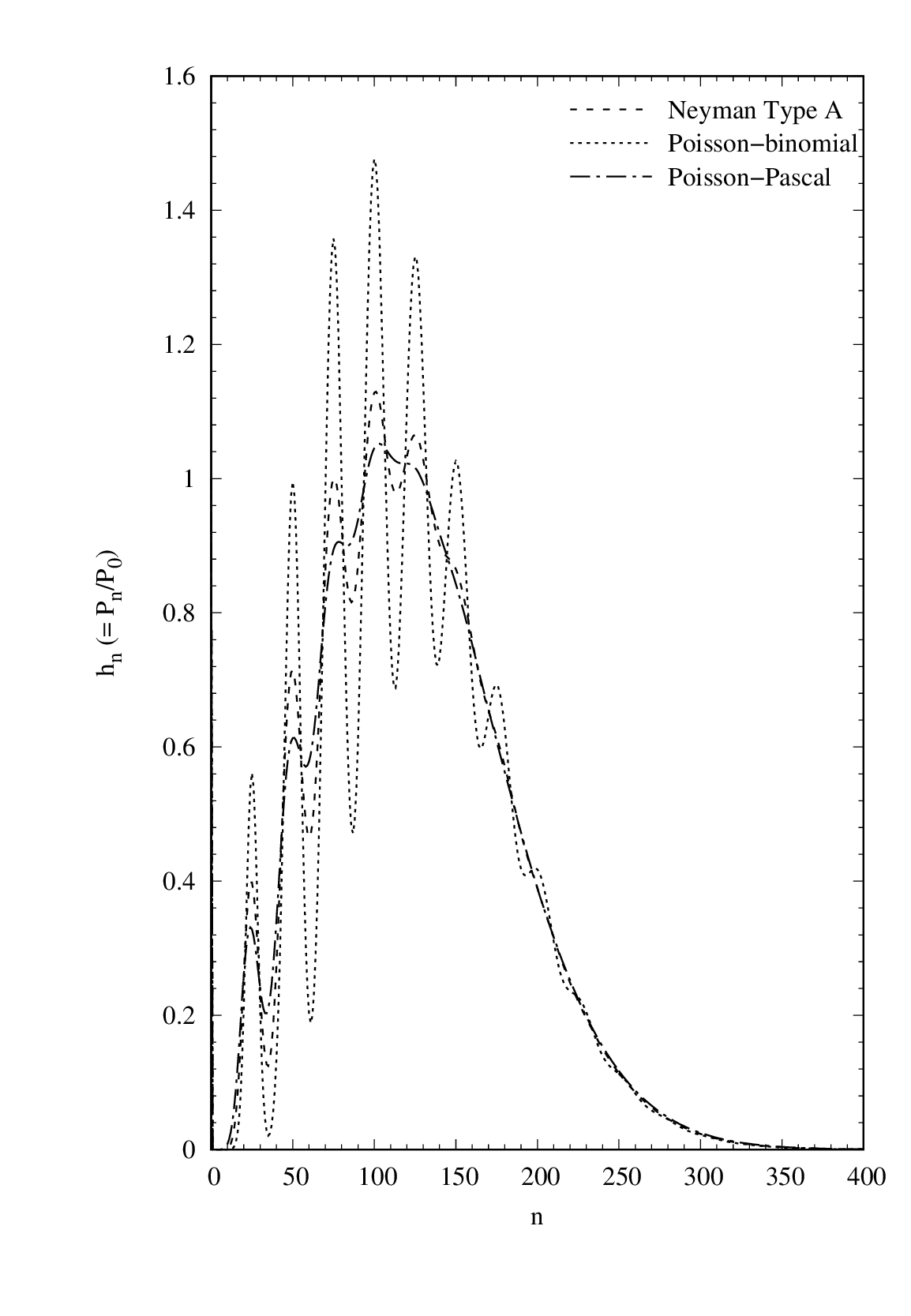}
\caption{\small
\label{fig:neyman_binom_pascal_pmf_lam5}
Plots of the scaled pmf $h_n(=P_n/P_0)$ of the
Neyman Type A (dash),
Poisson-binomial (dots) and
Poisson-Pascal (dotdash) distributions,
with the respective parameter values $(\lambda,\phi)=(5,25)$, $(\lambda,k,p)=(5,50,0.5)$ and $(\Lambda,k,P)=(5,50,0.5)$.
Note that $h_0=1$ in all cases, although this is difficult to observe in the plot.}
\end{figure}

\newpage
\begin{figure}[!htb]
\centering
\includegraphics[width=0.75\textwidth]{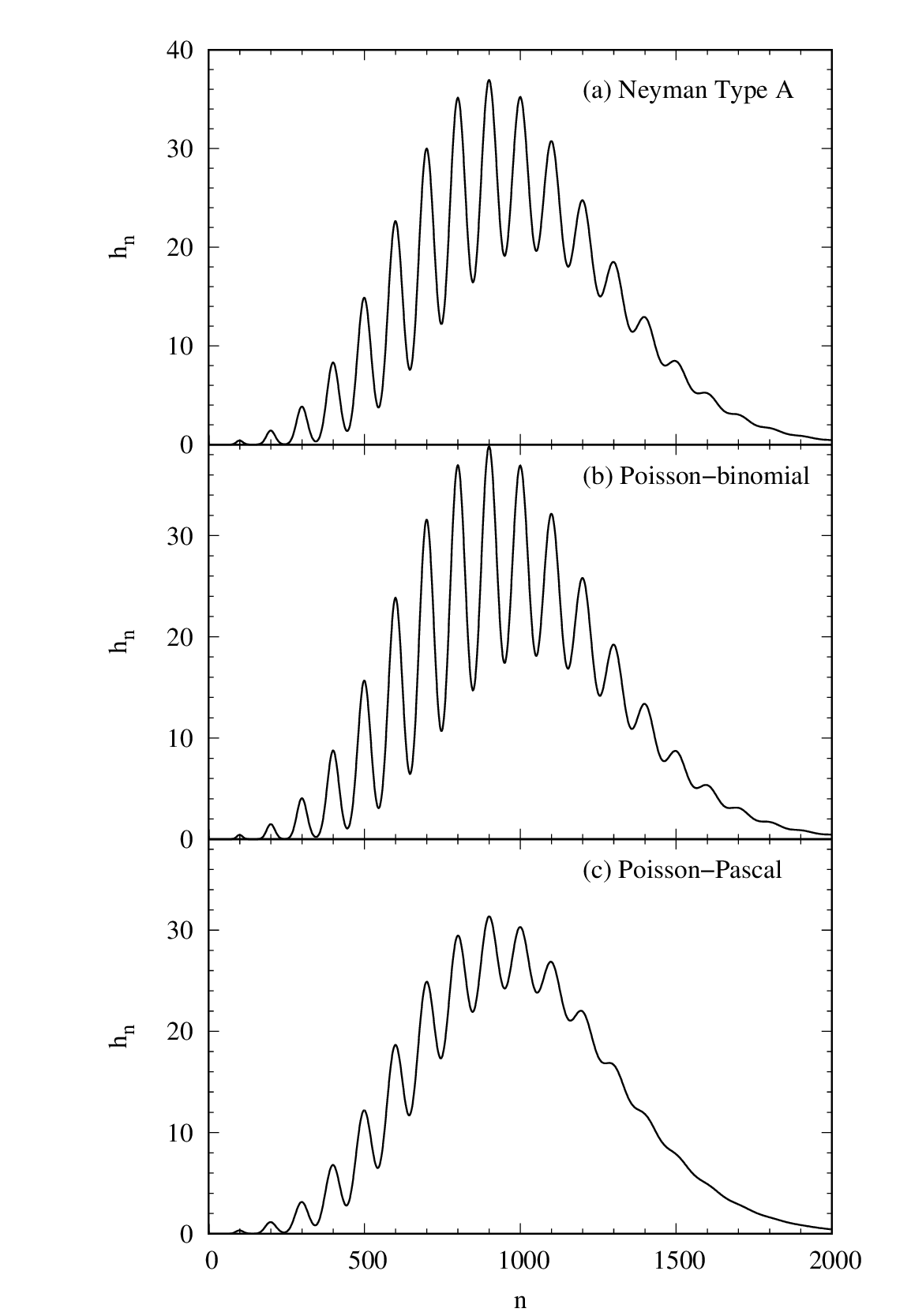}
\caption{\small
\label{fig:neyman_binom_pascal_pmf_lam10}
Plots of the scaled pmf $h_n(=P_n/P_0)$ of the
Neyman Type A (top),
Poisson-binomial (middle) and
Poisson-Pascal (bottom) distributions,
with the respective parameter values $(\lambda,\phi)=(10,100)$, $(\lambda,k,p)=(10,1000,0.1)$ and $(\Lambda,k,P)=(10,200,0.5)$.
Similar to Fig.~\ref{fig:neyman_binom_pascal_pmf_lam5}, but with different parameter values.}
\end{figure}

\newpage
\begin{figure}[!htb]
\centering
\includegraphics[width=0.75\textwidth]{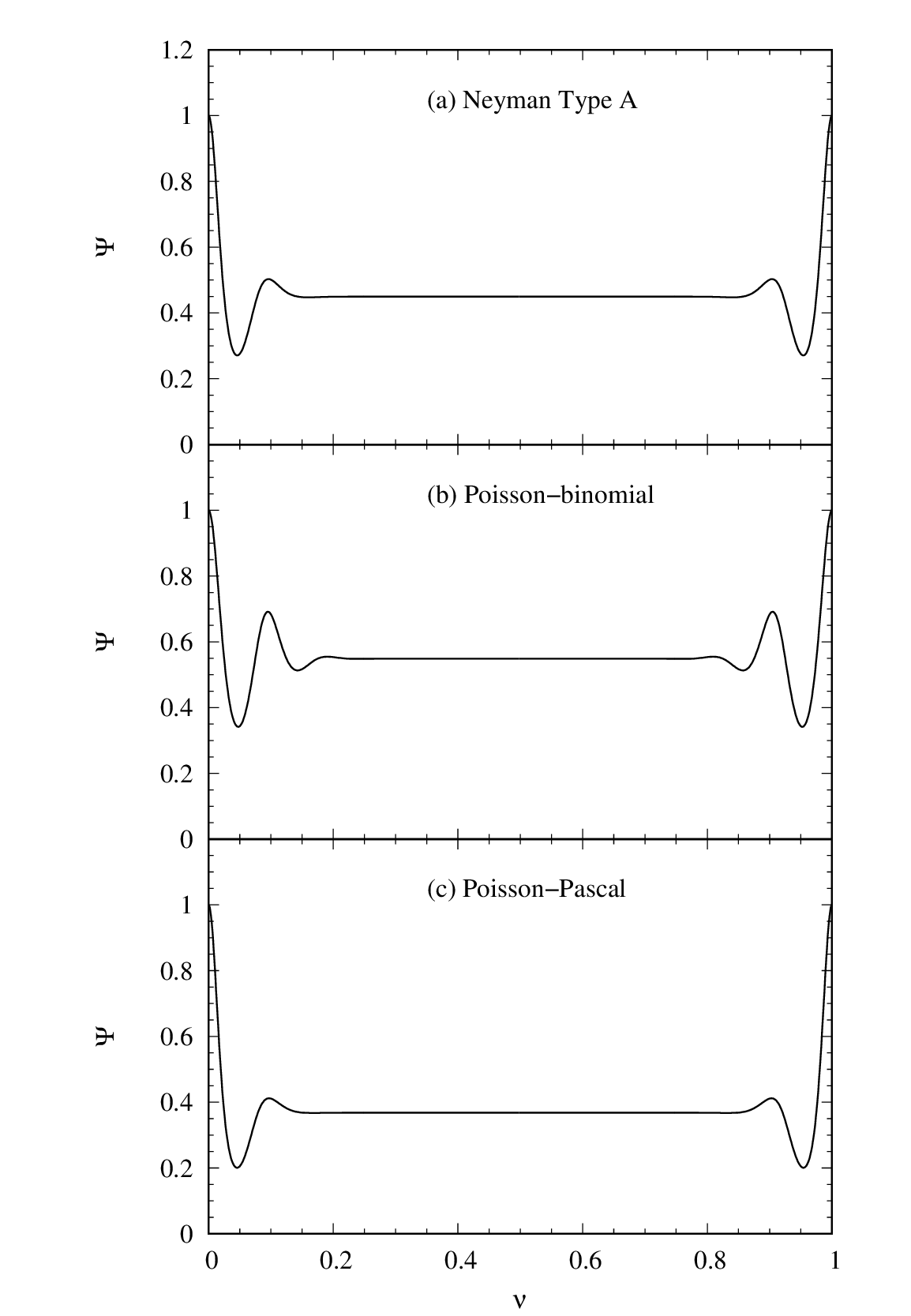}
\caption{\small
\label{fig:neyman_binom_pascal_ps}
Plots of the power spectra of Neyman Type A (top), Poisson-binomial (middle) and (c) Poisson-Pascal (bottom).
The respective parameter values are stated in the text.}
\end{figure}

\newpage
\begin{figure}[!htb]
\centering
\includegraphics[width=0.75\textwidth]{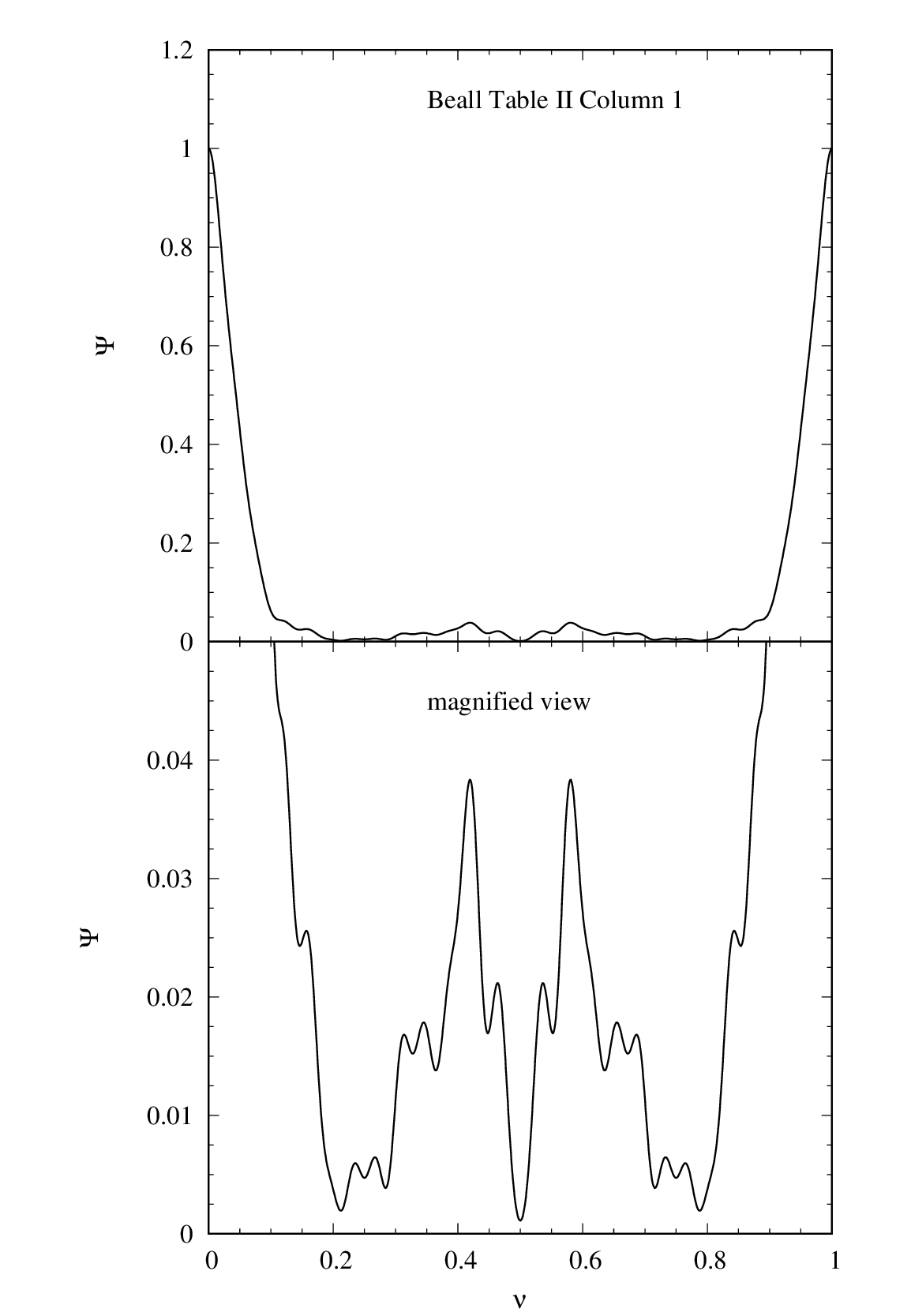}
\caption{\small
\label{fig:Beall_TableII_Col1_ps}
Graph of the power spectrum of the data in \cite{Beall} Table II column 1 (upper panel),
with a magnified view in the lower panel.}
\end{figure}

\newpage
\begin{figure}[!htb]
\centering
\includegraphics[width=0.75\textwidth]{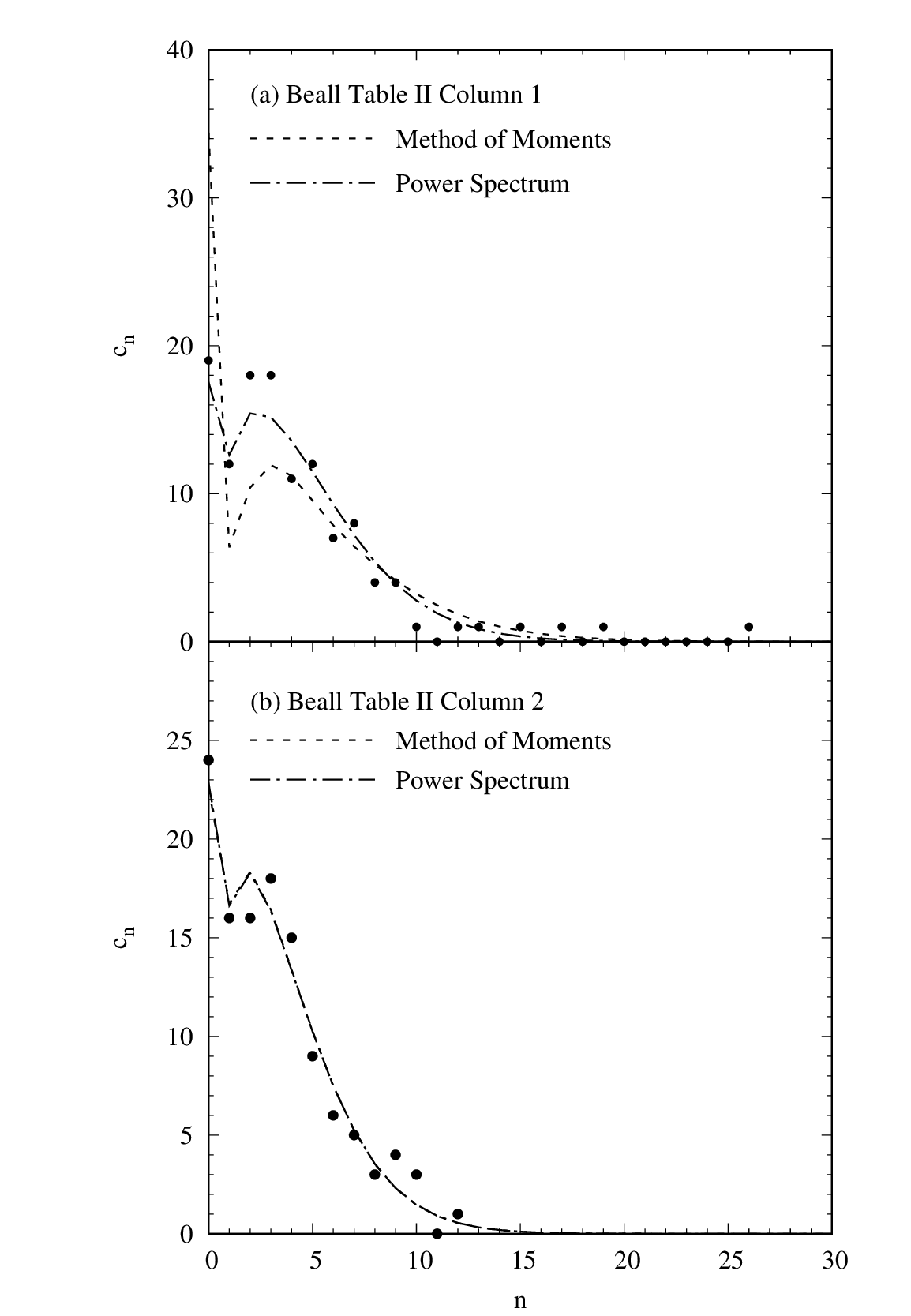}
\caption{\small
\label{fig:Beall_TableII_Col12_fits}
Plot of the data in \cite{Beall} Table II columns 1 (top) and 2 (bottom).
Here $c_n$ denotes the count in each histogram bin.
In each panel, the points indicate the data and the curves are the fits using the method of moments (dash) and power spectrum (dotdash).}
\end{figure}

\newpage
\begin{figure}[!htb]
\centering
\includegraphics[width=0.75\textwidth]{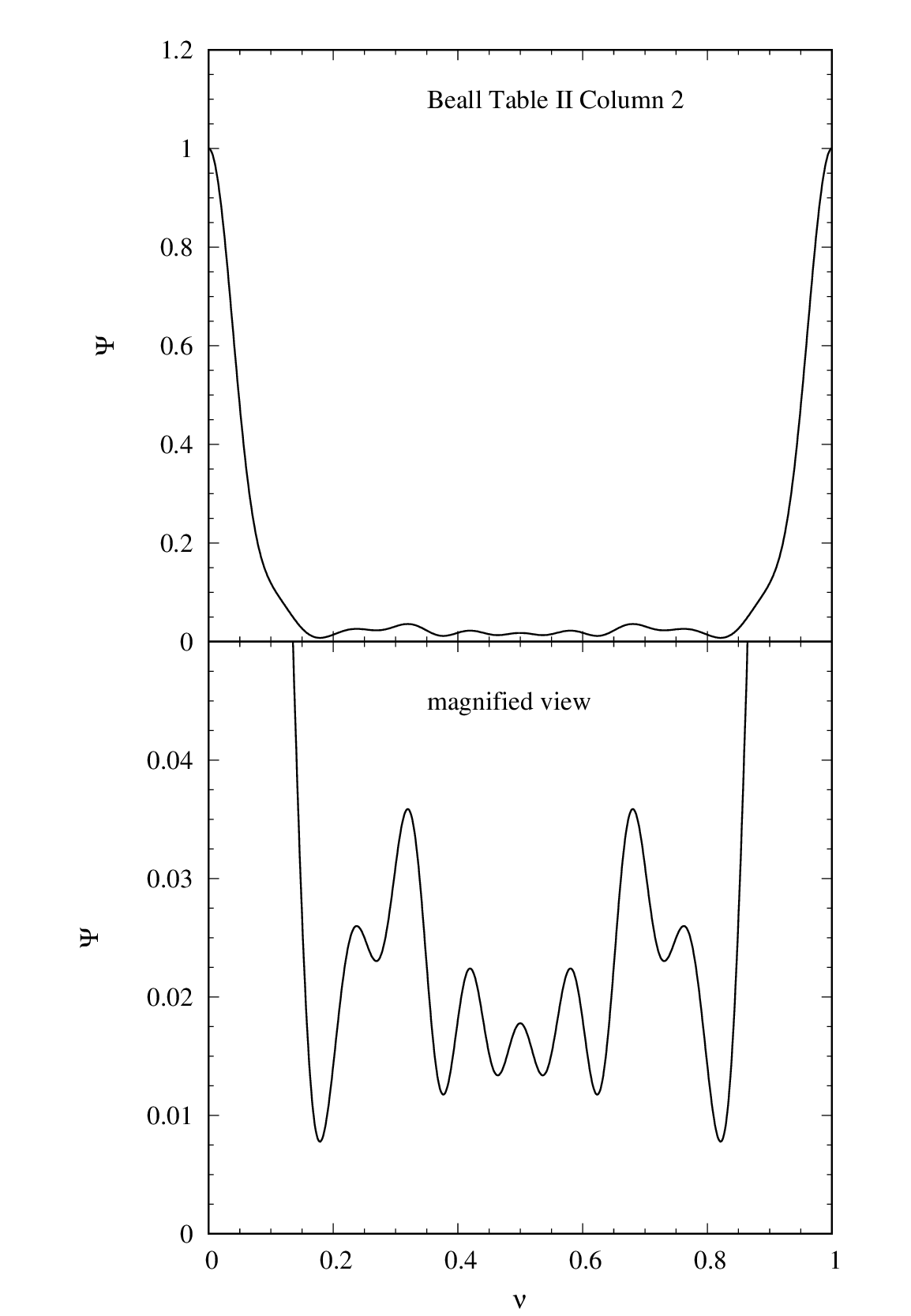}
\caption{\small
\label{fig:Beall_TableII_Col2_ps}
Graph of the power spectrum of the data in \cite{Beall} Table II column 2 (upper panel),
with a magnified view in the lower panel.}
\end{figure}

\newpage
\begin{figure}[!htb]
\centering
\includegraphics[width=0.75\textwidth]{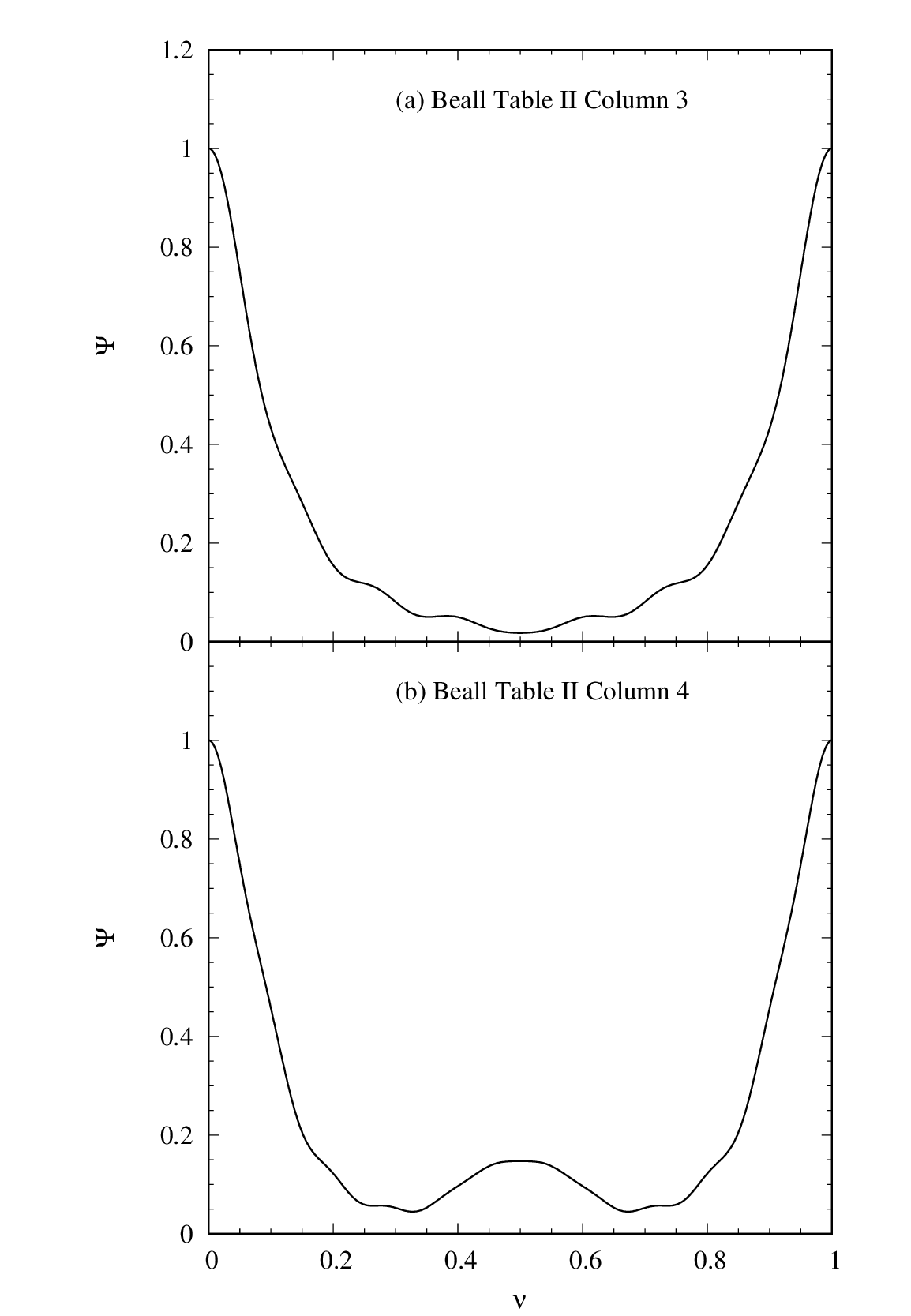}
\caption{\small
\label{fig:Beall_TableII_Col34_ps}
Graphs of the power spectra of the data in \cite{Beall} Table II columns 3 (top) and 4 (bottom).}
\end{figure}

\newpage
\begin{figure}[!htb]
\centering
\includegraphics[width=0.75\textwidth]{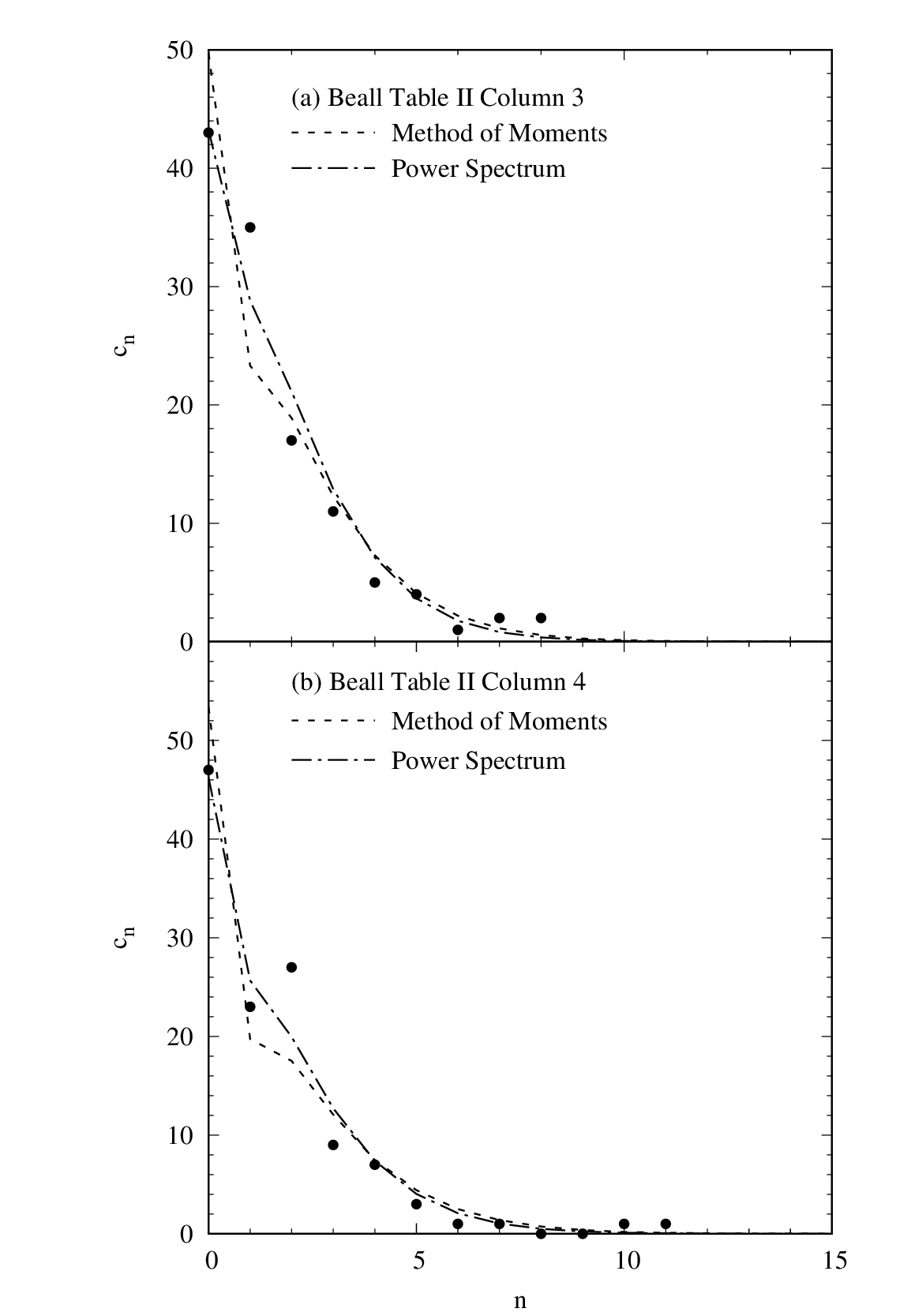}
\caption{\small
\label{fig:Beall_TableII_Col34_fits}
Plot of the data in \cite{Beall} Table II columns 3 (top) and 4 (bottom).
Here $c_n$ denotes the count in each histogram bin.
In each panel, the points indicate the data and the curves are the fits using the method of moments (dash) and power spectrum (dotdash).}
\end{figure}

\newpage
\begin{figure}[!htb]
\centering
\includegraphics[width=0.75\textwidth]{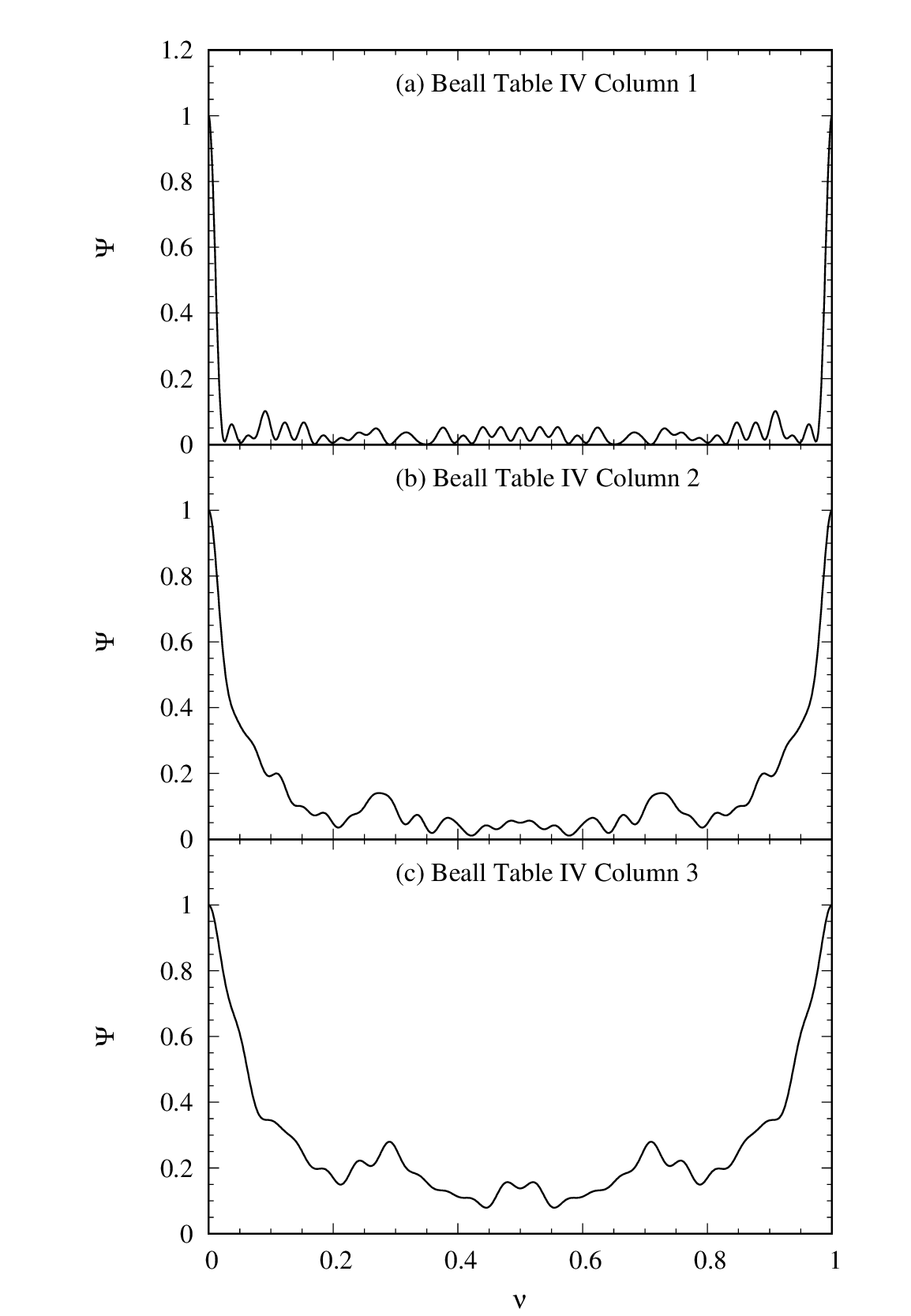}
\caption{\small
\label{fig:Beall_TableIV_Col123_ps}
Graph of the power spectra of the data in \cite{Beall} Table IV column 1 (top), 2 (middle) and 3 (bottom).}
\end{figure}

\newpage
\begin{figure}[!htb]
\centering
\includegraphics[width=0.75\textwidth]{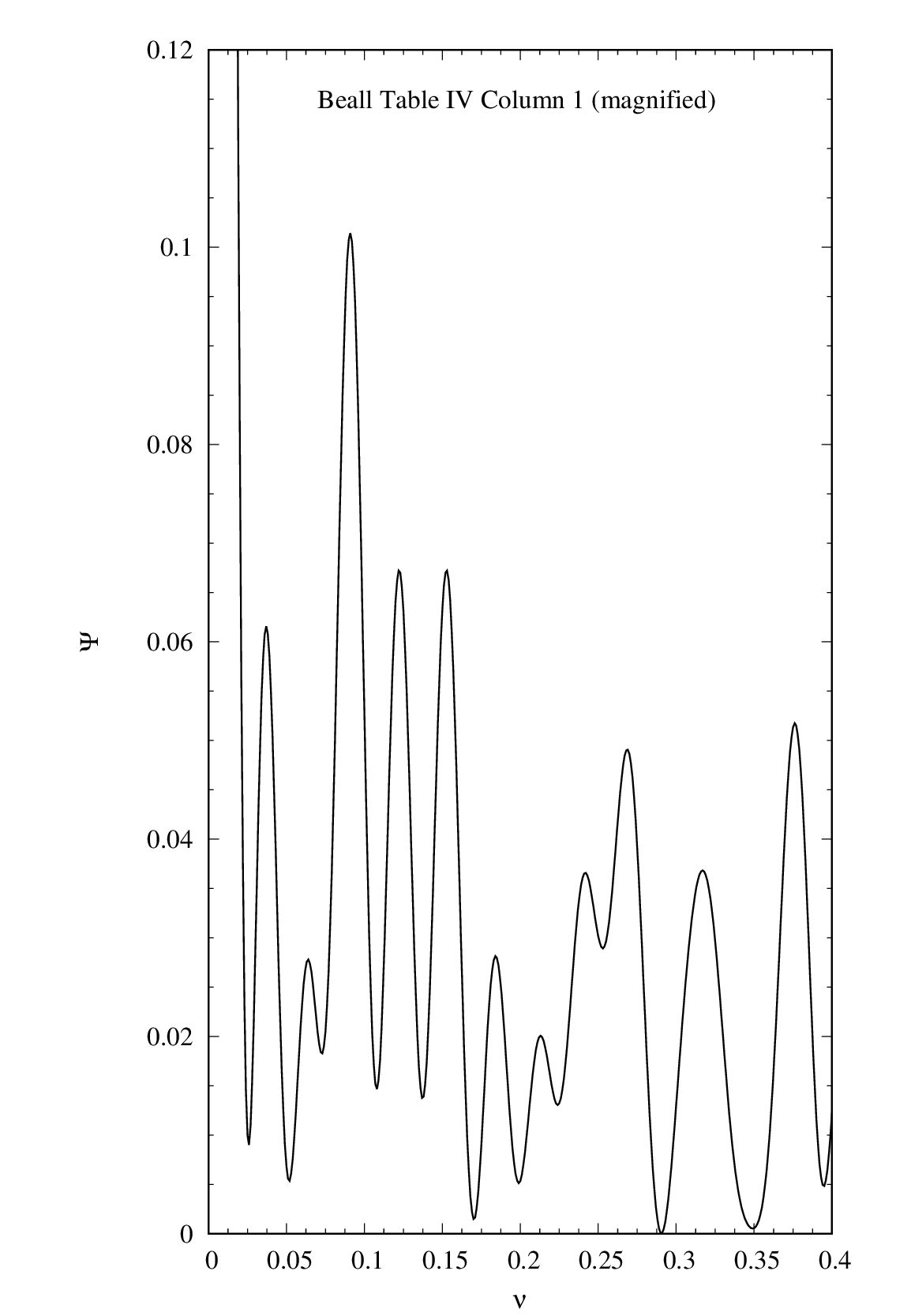}
\caption{\small
\label{fig:Beall_TableIV_Col1_mag_ps}
Magnified view of the power spectrum of the data in \cite{Beall} Table IV column 1
(top panel of Fig.~\ref{fig:Beall_TableIV_Col123_ps}).}
\end{figure}

\newpage
\begin{figure}[!htb]
\centering
\includegraphics[width=0.75\textwidth]{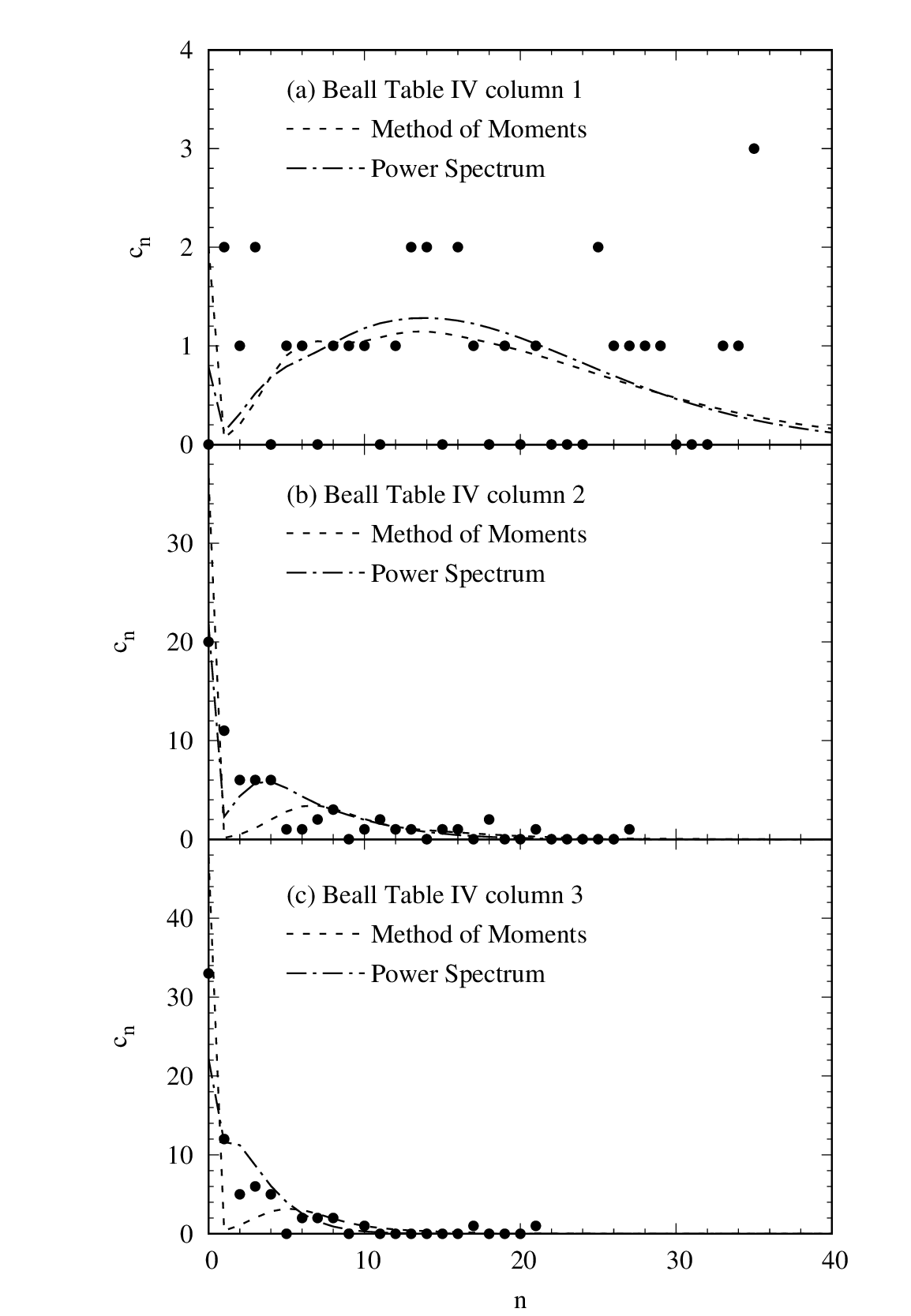}
\caption{\small
\label{fig:Beall_TableIV_Col123_fits}
Plot of the data in \cite{Beall} Table II columns 1 (top), 2 (middle) and 3 (bottom).
Here $c_n$ denotes the count in each histogram bin.
In each panel, the points indicate the data and the curves are the fits using the method of moments (dash) and power spectrum (dotdash).
Also $c_n$ denotes the count of the data in each bin.}
\end{figure}

\newpage
\begin{figure}[!htb]
\centering
\includegraphics[width=0.75\textwidth]{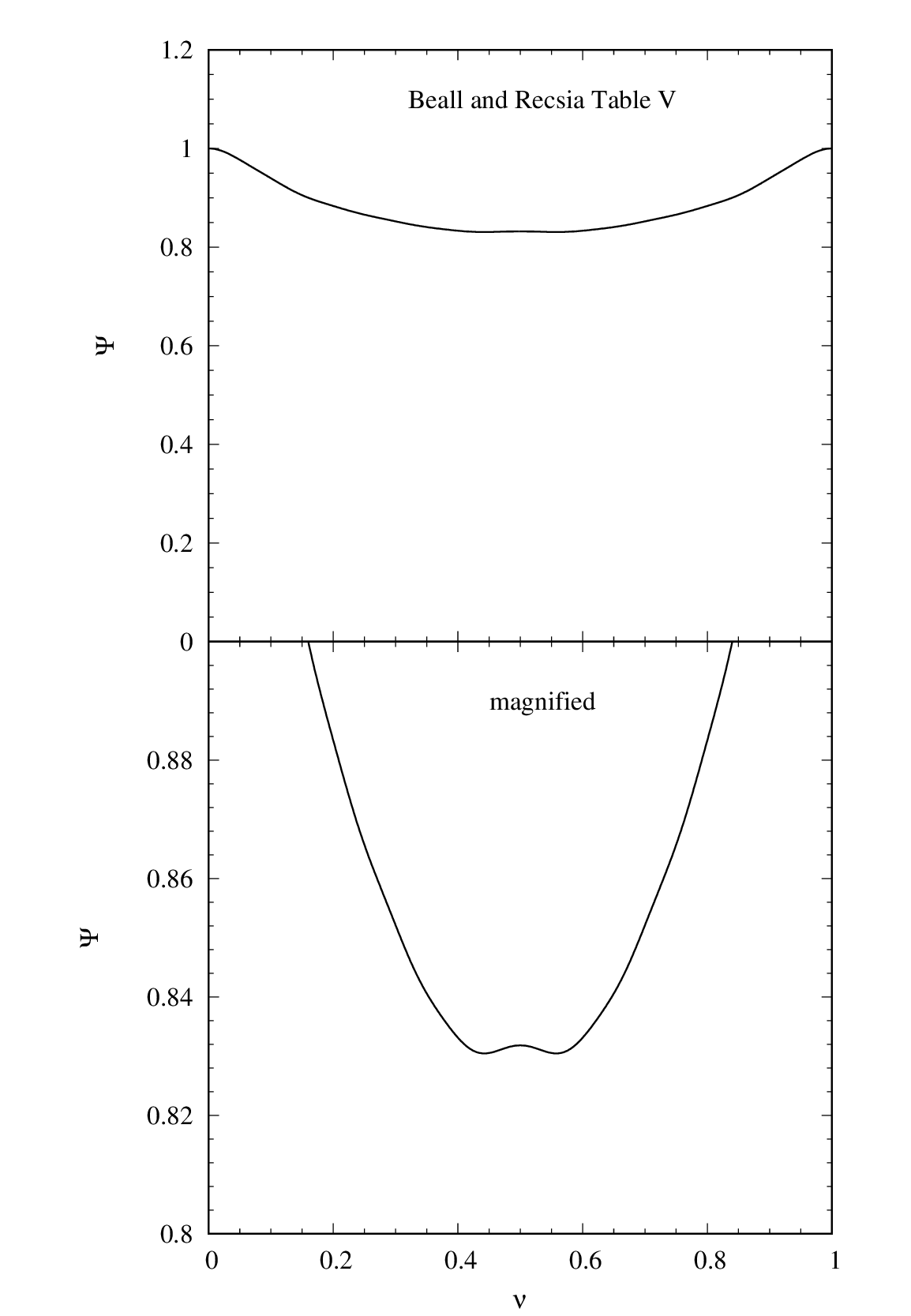}
\caption{\small
\label{fig:BR_Table5_PS1000}
Graph of the power spectrum (and magnified view) of the data by \cite{BeallRescia1953}, Table V.}
\end{figure}

\newpage
\begin{figure}[!htb]
\centering
\includegraphics[width=0.75\textwidth]{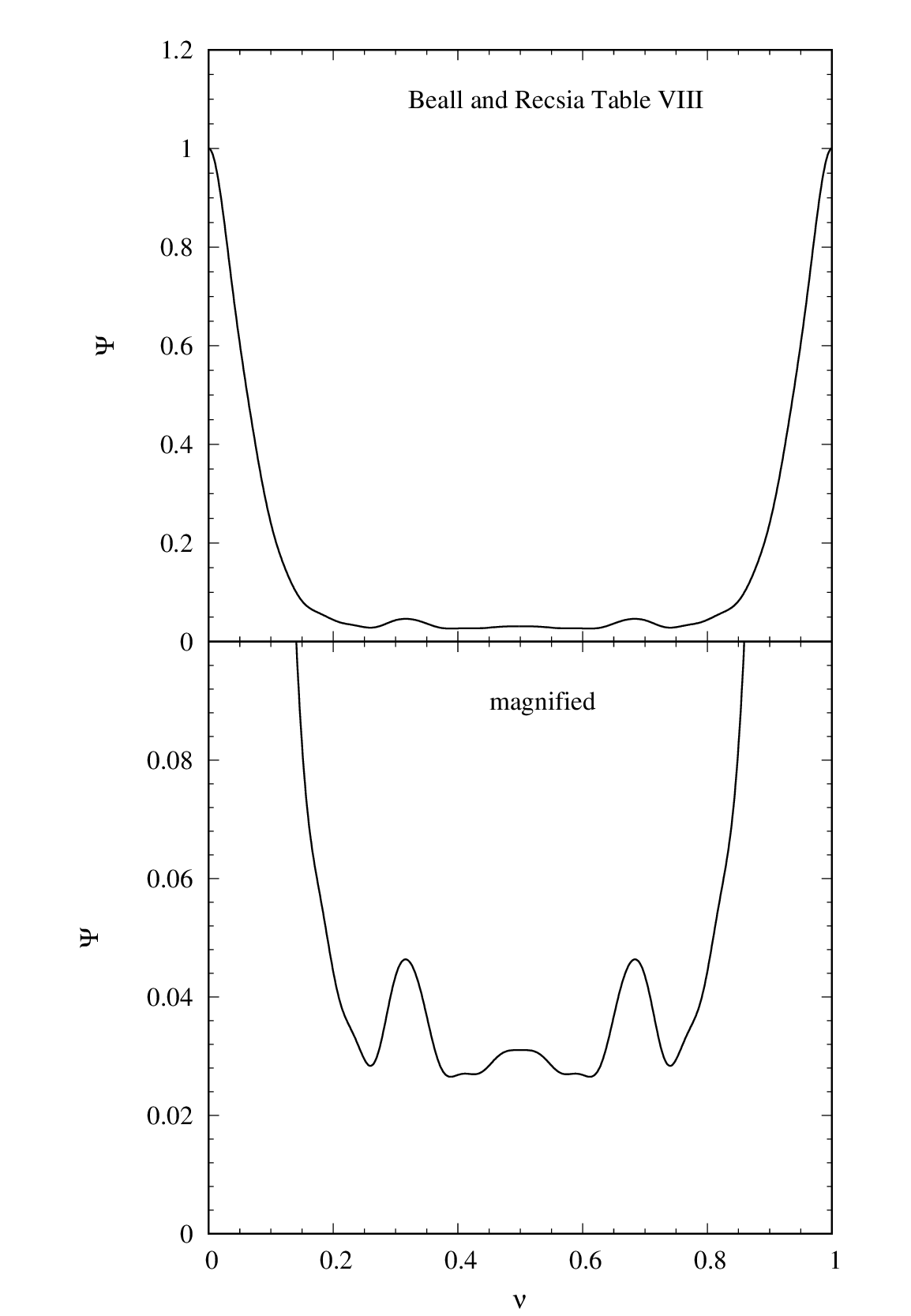}
\caption{\small
\label{fig:BR_Table8_PS1000}
Graph of the power spectrum (and magnified view) of the data by \cite{BeallRescia1953}, Table VIII.}
\end{figure}

\newpage
\begin{figure}[!htb]
\centering
\includegraphics[width=0.75\textwidth]{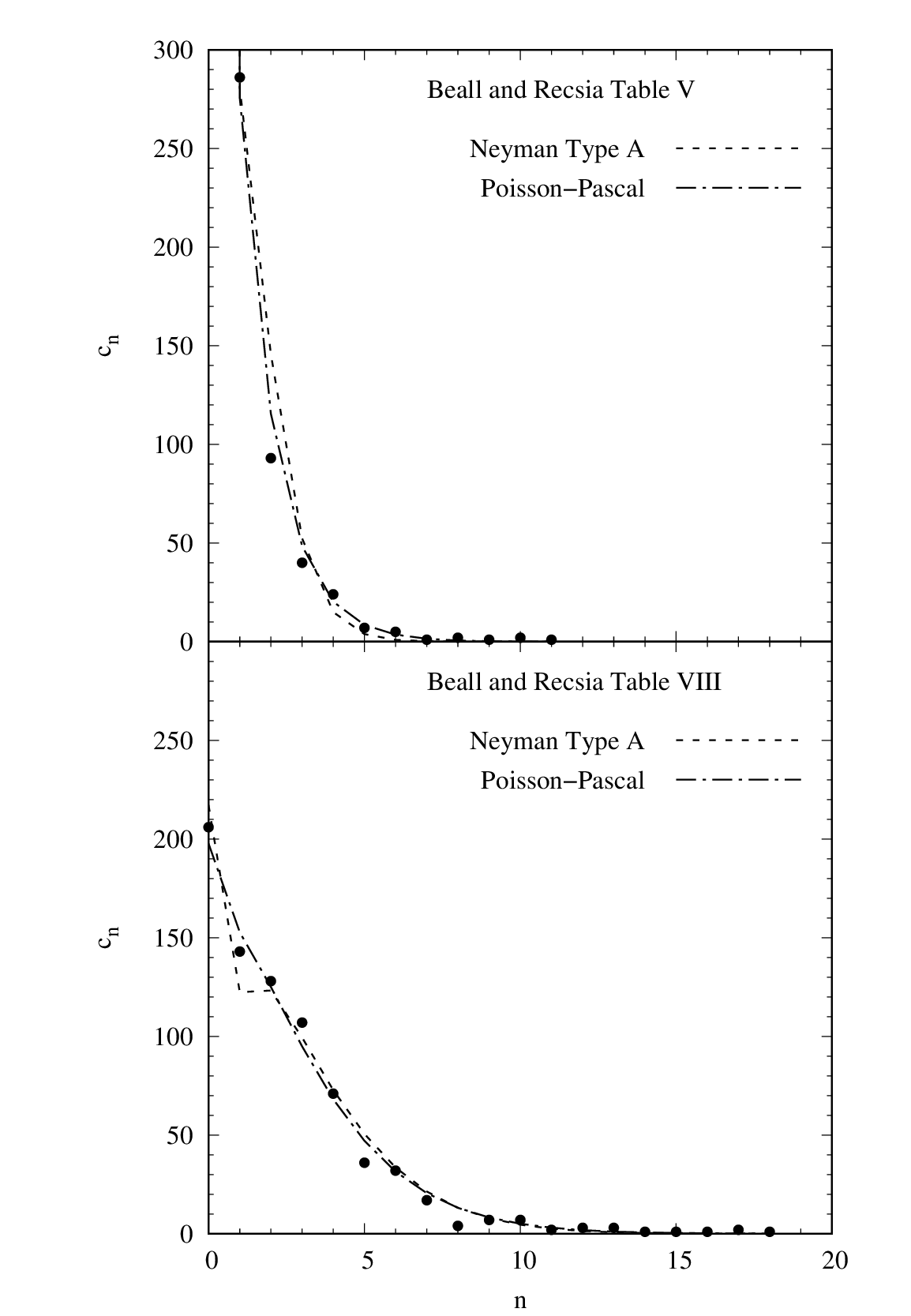}
\caption{\small
\label{fig:BR_Table58_fits}
Graphs of the data in \cite{BeallRescia1953}
for Table V (top) and VIII (bottom).
Here $c_n$ denotes the count in each histogram bin.
The curves are fits using the Neyman Type A distribution (dash) and Poisson-Pascal (dotdash).
For Table V, the point at $n=0$ has been excluded because it is much taller that the rest.}
\end{figure}

\newpage
\begin{figure}[!htb]
\centering
\includegraphics[width=0.75\textwidth]{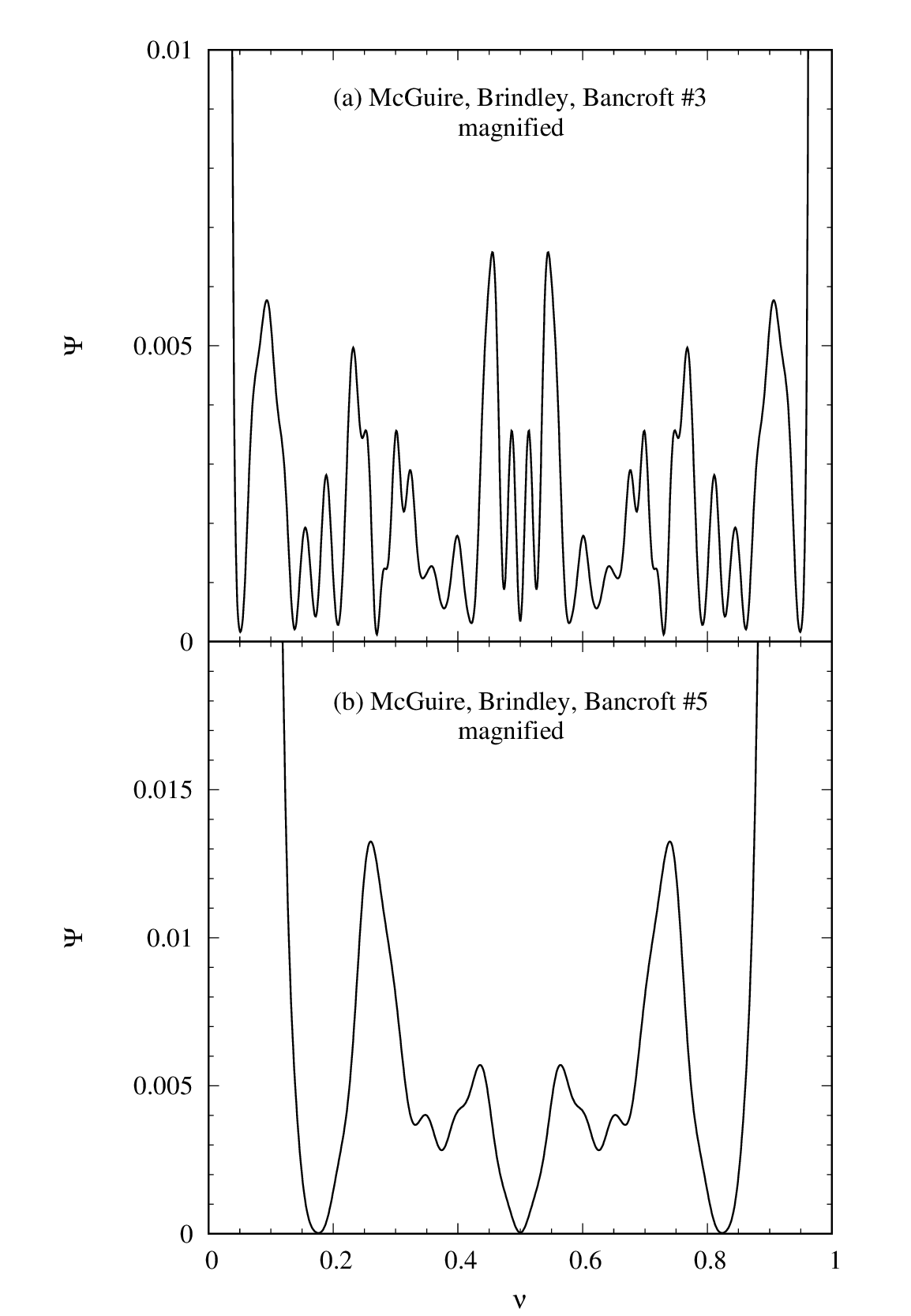}
\caption{\small
\label{fig:McGBB_dist35_PS1000}
Graphs of the power spectra (magnified views) of the data in 
\cite{McGBB} for distribution 3 (top) and 5 (bottom).}
\end{figure}

\newpage
\begin{figure}[!htb]
\centering
\includegraphics[width=0.75\textwidth]{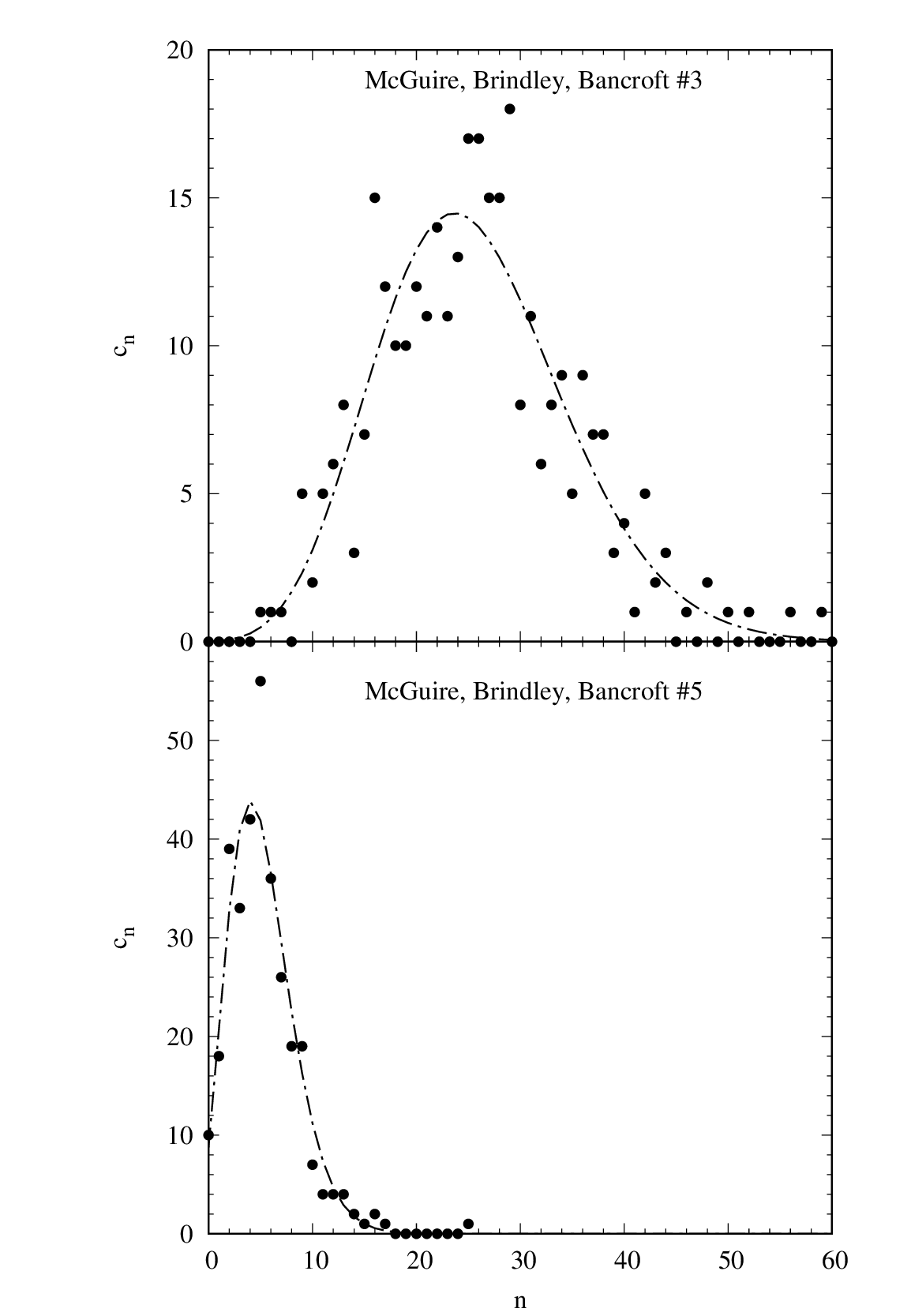}
\caption{\small
\label{fig:McGBB_dist35_hist}
Graphs of the data in \cite{McGBB}
for distributions 3 (top) and 5 (bottom).
Here $c_n$ denotes the count in each histogram bin.
The curves are fits using the Neyman Type A distribution.}
\end{figure}

\newpage
\begin{figure}[!htb]
\centering
\includegraphics[width=0.75\textwidth]{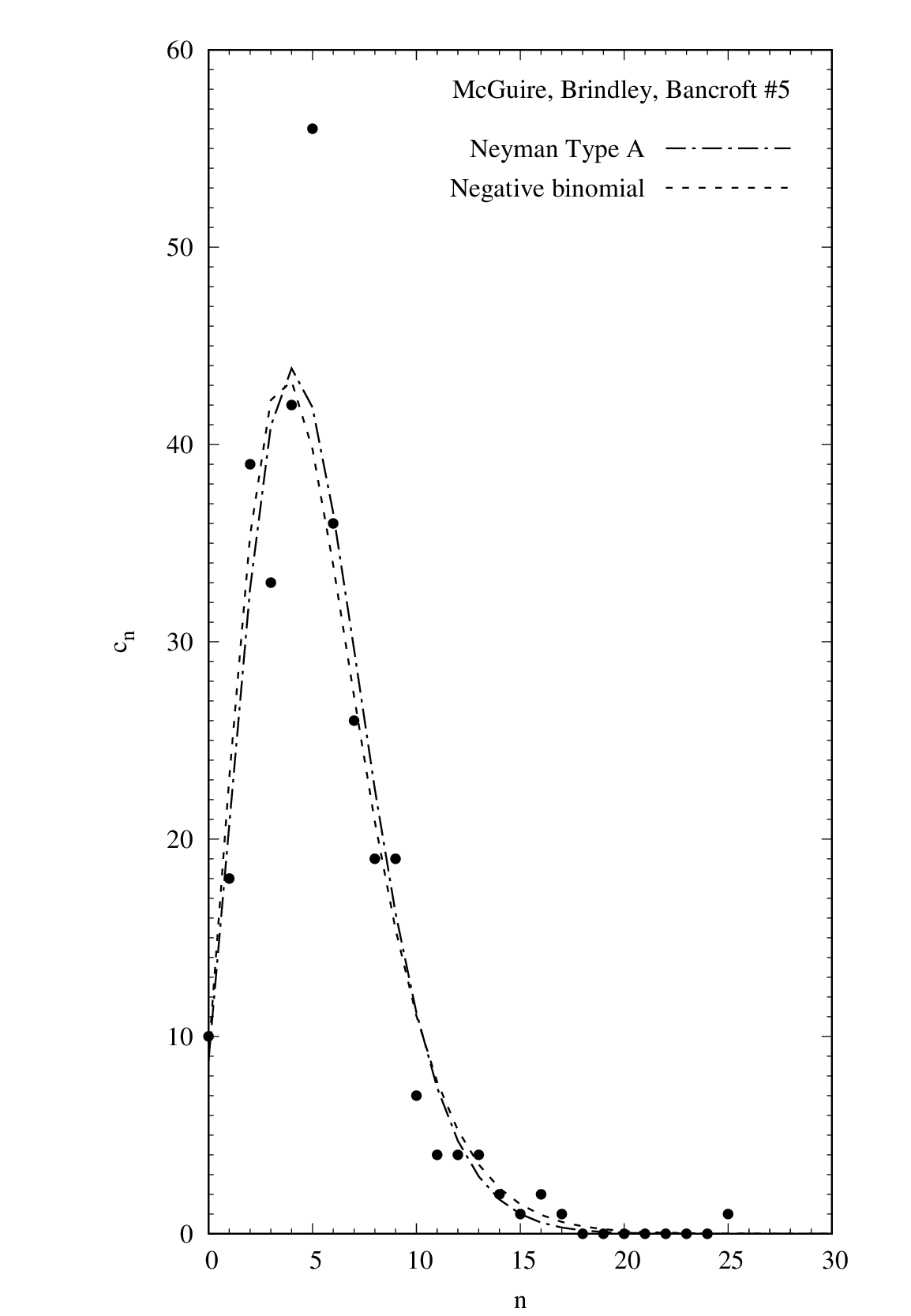}
\caption{\small
\label{fig:McGBB_dist5_hist}
Graph of the data in \cite{McGBB}, distribution 5.
Here $c_n$ denotes the count in each histogram bin.
The dotdash and dashed curves are fits using the Neyman Type A
and negative binomial distribution, respectively.}
\end{figure}

\newpage
\begin{figure}[!htb]
\centering
\includegraphics[width=0.75\textwidth]{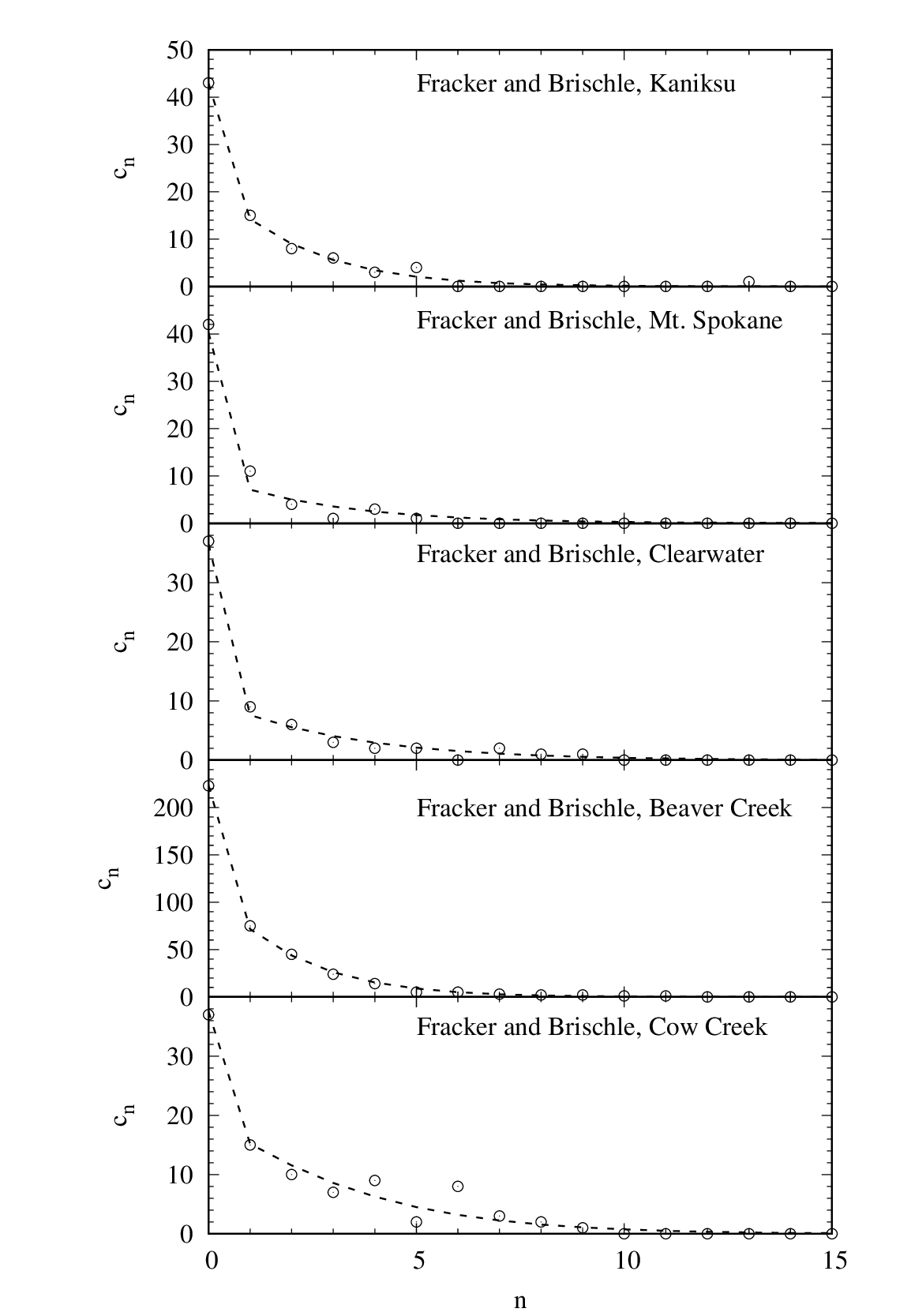}
\caption{\small
\label{fig:FB_hist}
Graph of the data (points) for the five datasets in \cite{FrackerBrischle1944}.
Here $c_n$ denotes the count in each histogram bin.
The fits (dashed curves) were calculated using the geometric Poisson distribution.}
\end{figure}

\setcounter{section}{0}
\vfill\pagebreak
\begin{table}[!htb]
\begin{tabular}{rrrrrrrrrrrrrrrrrr}
  \hline
  Column & $\bar{x}$ & $s^2$ & $m_1$ & $m_2$ & $\Delta_{\rm MM}$ \\
  \hline
  1 & $1.4000$ & $2.3272$ & $2.1140$ & $0.6623$ & $0.1517$ \\
  2 & $0.5046$ & $0.5841$ & $3.2043$ & $0.1575$ & $0.2021$ \\
  3 & $0.8523$ & $1.1386$ & $2.5372$ & $0.3359$ & $0.3081$ \\
  4 & $0.4123$ & $0.5208$ & $1.5664$ & $0.2632$ & $0.0266$ \\
\hline
\end{tabular}
\caption{\label{tb:Beall_TableI_fit}
Parameter estimation and goodness of fit values for the data in the columns in \cite{Beall}, Table I.}
\end{table}

\vfill\pagebreak
\begin{table}[!htb]
\begin{tabular}{rrrrrrrrrrrrrrrrrr}
  \hline
  Column & $N_c$ & $\bar{x}$ & $s^2$ & $m_1$ & $m_2$ & $\Delta_{\rm MM}$ & $\hat{\lambda}$ & $\hat{\phi}$ & $\Delta_{\rm PS}$  \\
  \hline
  1  &  120 & $4.0333$ & $16.4527$
  & $1.3099$ & $3.0792$ & $0.197$ 
  & $2.3393$ & $1.7241$ & $0.021$ \\
  2 &  120 & $3.1667$ & $7.7703$
  & $2.1782$ & $1.4538$ & $0.025$ 
  & $2.1533$ & $1.4706$ & $0.024$ \\  
  3 &  120 & $1.4833$ & $3.1930$
  & $1.2870$ & $1.1526$ & $0.514$ 
  & $1.8541$ & $0.8$ & $0.178$ \\
  4 &  120 & $1.5083$ & $3.6302$
  & $1.0722$ & $1.4068$ & $0.362$ 
  & $1.5083$ & $1$ & $0.173$ \\
\hline
\end{tabular}
\caption{\label{tb:Beall_Table2}
Count total, sample mean and variance and parameter estimates for the data in the columns in \cite{Beall}, Table II.}
\end{table}

\vfill\pagebreak
\begin{table}[!htb]
\begin{tabular}{rrrrrrrrrrrrrrrrrr}
  \hline
  Column & $N_c$ & $\bar{x}$ & $s^2$ & $m_1$ & $m_2$ & $\Delta_{\rm MM}$ & $\hat{\lambda}$ & $\hat{\phi}$ & $\Delta_{\rm PS}$  \\
  \hline
  1 & 31 & $17.2581$ & $121.7399$
  & $2.7441$ & $6.2892$ & $0.0083$ 
  & $3.7105$ & $4.6512$ & $0.0075$ \\
  2 & 67 & $4.2687$ & $33.8681$
  & $0.6051$ & $7.0544$ & $0.2119$ 
  & $1.1525$ & $3.7037$ & $0.0548$ \\
  3 & 70 & $2.1429$ & $13.8939$
  & $0.3842$ & $5.5778$ & $0.4098$ 
  & $1.5214$ & $1.4085$ & $0.1875$ \\
\hline
\end{tabular}
\caption{\label{tb:Beall_Table4}
Count total, sample mean and variance and parameter estimates for the data in the columns in \cite{Beall}, Table IV.}
\end{table}

\vfill\pagebreak
\begin{table}[!htb]
\begin{tabular}{ccccc}
  \hline
  $\nu_{\rm peak}$ & \qquad & $\Delta (\times10^3)$ \\
  \hline
$0.093$ & \qquad & $17.74$ \\
$0.125$ & \qquad & $10.79$ \\
$0.154$ & \qquad & $8.49$ \\
$0.185$ & \qquad & $7.69$ \\
$0.215$ & \qquad & $7.52$ \\
$0.244$ & \qquad & $7.59$ \\
$0.270$ & \qquad & $7.73$ \\
$0.320$ & \qquad & $8.09$ \\
$0.380$ & \qquad & $8.52$ \\
\hline
\end{tabular}
\caption{\label{tb:Beall_TableIV_Col1_Delta}
  Values of $\Delta$ for various peaks in the power spectrum for the data in \cite{Beall}, Table IV Column 1.}
\end{table}

\vfill\pagebreak
\begin{table}[!htb]
\begin{tabular}{llrrrrrrrrrr}
  \hline
  Distribution & Method & $\Delta$ (Table V) & $\Delta$ (Table VIII) \\ \hline
  NTA & MM  
  & $14.42$
  & $1.42$ \\
  NTA & PS
  & $1.97$
  & $0.18$ \\
  geom & MM
  & $2.86$ 
  & $0.28$ \\
  geom & alt
  & $0.50$
  & $0.10$ \\
  PP & PS
  & $0.38$
  & $0.10$ \\
  \hline
\end{tabular}
\caption{\label{tb:BR_Delta}
Goodness of fit for the data in \cite{BeallRescia1953}, Tables V and VIII.
The acronyms are explained in the text.}
\end{table}

\vfill\pagebreak
\begin{table}[!htb]
\begin{tabular}{llrrrrrrrrrr}
  \hline
  Distribution & Method & $\Delta (\times10^3)$ (dist.~3) & $\Delta (\times10^3)$ (dist.~5) \\ \hline
  PB & MM
  & $8.29$
  & $135.6$ \\
  NTA & MM
  & $8.40$
  & $134.7$ \\
  PB & PS
  & $9.28$
  & $111.9$ \\
  NTA & PS
  & $8.40$
  & $109.8$ \\
  NB & MM
  & $8.99$
  & $128.5$ \\
  \hline
\end{tabular}
\caption{\label{tb:McGBB_Delta}
Goodness of fit for the data in \cite{McGBB}, Distributions 3 and 5.
The acronyms are explained in the text.}
\end{table}

\vfill\pagebreak
\begin{table}[!htb]
\begin{tabular}{lrrrrrrrrrr}
  \hline
  Dataset & $\hat{\lambda}$ & $\hat{p}$ \\ \hline
  Kaniksu & 0.6208 & 0.4716 \\
  Mt.~Spokane & 0.4212 & 0.6544 \\
  Clearwater & 0.5480 & 0.6385 \\
  Beaver Creek & 0.5843 & 0.4488 \\
  Cow Creek & 0.9324 & 0.5528 \\
  \hline
\end{tabular}
\caption{\label{tb:FB_fits}
Values of estimators $(\hat{\lambda},\hat{p})$ for the geometric Poisson distribution,
to fit the datasets of \cite{FrackerBrischle1944}.}
\end{table}


\begin{thebibliography}{refs}
\bibitem[Anscombe (1950)]{Anscombe}
  Anscombe, F.J.: Sampling Theory of the Negative Binomial and Logarithmic Series Distributions.
  Biometrika \textbf{37}, 358--382 (1950)

\bibitem[Beall (1940)]{Beall}
Beall, G.: The Fit and Significance of Contagious Distributions when Applied to Observations on Larval Insects.
Ecology \textbf{21}, 460--474 (1940)

\bibitem[Beall and Recsia (1953)]{BeallRescia1953}
Beall, G. and Rescia, R.R.: A Generalization of Neyman's Contagious Distributions.
Biometrics 9:354--386 (1953)

\bibitem[Bliss and Fisher (1953)]{BlissFisher1953}
Bliss, C.I. and Fisher R.A.: Fitting the Negative Binomial Distribution to Biological Data.
Biometrics \textbf{14}, 176--200 (1953)

\bibitem[Fracker and Brischle (1944)]{FrackerBrischle1944}
Fracker, S.B. and Brischle, H.A.: Measuring the Local Distribution of Ribes.
Ecology \textbf{25}, 283--303 (1944)

\bibitem[Gurland (1957)]{Gurland1957}
Gurland, J.: Some Interrelations among Compound and Generalized Distributions.
Biometrika \textbf{44}, 265--268 (1957)

\bibitem[Gurland (1958)]{Gurland1958}
Gurland, J.: A Generalized Class of Contagious Distributions.
Biometrics \textbf{14}, 229--249 (1958)

\bibitem[Johnson, Kemp and Kotz (2005)]{JohnsonKempKotz}
Johnson, N.L., Kemp, A.W. and Kotz, S.:
Univariate Discrete Distributions, 3rd ed.~Wiley-Interscience, Hoboken, New Jersey, USA (2005)

\bibitem[McGuire, Brindley and Bancroft (1957)]{McGBB}
McGuire, J.U., Brindley, T.A. and Bancroft, T.A.: The Distribution of European Corn Borer Larvae Pyrausta nubilalis (Hbn.), in Field Corn.
Biometrics \textbf{13}, 65--78 (1957)
\end{thebibliography}
\end{document}